\documentclass[11pt,a4paper]{article}
\pdfoutput=1
\usepackage[utf8x]{inputenc} 
\usepackage{jheppub}
\usepackage{amsfonts}
\usepackage{amsmath}
\usepackage{slashed}
\usepackage{lscape}
\usepackage{amssymb}
\usepackage{graphicx}
\usepackage{epic}
\usepackage{eepic}
\usepackage{epsfig}
\usepackage{latexsym}
\usepackage{color}
\usepackage{float}
\usepackage{multirow}
\usepackage{hyperref}
\usepackage{enumitem}
\hypersetup{colorlinks=true,citecolor=blue,linkcolor=magenta,urlcolor=blue}
\usepackage[caption=false]{subfig}
\usepackage{natbib}
\usepackage{relsize}
\usepackage{breqn}
\usepackage{shorthand}
\usepackage{subfig}
\usepackage[]{appendix}
\usepackage[normalem]{ulem}

\definecolor{darkred}{cmyk}{0,1,1,0.4}

\usepackage{color}
 \usepackage[normalem]{ulem}
\long\def\/*#1*/{}
\usepackage{color}
 \usepackage[normalem]{ulem}
\definecolor{darkgreen}{cmyk}{1,0,1,0.4}
\definecolor{darkred}{cmyk}{0,1,1,0.4}

\title{Flavored leptogenesis and Dirac CP violation}

\author[a]{Ananya Mukherjee,}
\emailAdd{ananyatezpur@gmail.com}

\author[b]{Nimmala Narendra,}
\emailAdd{nimmalanarendra@gmail.com}

\affiliation[a]{Department of Physics, University of Calcutta, 92 Acharya Prafulla Chandra Road, Kolkata 700009, India}
\affiliation[b]{Theoretical Physics Division, Physical Research Laboratory, Ahmedabad-380009, Gujarat, India}

\abstract{ In this work we pay special attention in establishing the crucial role of the Casas-Ibarra (CI) parameterisation in presence of two different orthogonal matrices, $R= \textbf{O} \,\rm e^{i {\bf A}}$ and $R= \textbf{O} \,\rm e^{{\bf A}}$ in order to investigate the role of Dirac CP violation in flavored leptogenesis. In the light of these two choices of the $R$ matrix we examine the connection between the low energy and high energy CP violations along with certain interesting predictions on the low energy parameters namely, the lightest neutrino mass and the Dirac CP phase ($\delta$). Considering the right handed neutrino (RHN) mass window to be $10^8$ GeV, we show that Dirac phase leptogenesis is possible with the choices of these two orthogonal matrices. The above forms of $R$ matrices allow us to choose a nearly degenerate spectrum for the RHN masses. The complex $R$ matrix predicts a maximal Dirac CP violation ($\delta\,=\,\pi/2$) for leptogenesis which, can be verified by the ongoing and upcoming searches for a precise $\delta$ measurement at the neutrino factories. We also discuss the phenomenological implications of these two case studies in the context of LFV considering the $\mu \rightarrow e\gamma$ decay process, interms of an indirect probe of the $R$ matrix parameter space. We report an upper bound on the lightest neutrino mass ($m_1$) of around $0.015$\,eV from the baryon asymmetry criteria for normal hierarchy of neutrino masses.}

\begin{document}

\maketitle
\def\lapp{\mathrel{\rlap{\raise.5ex\hbox{$<$}}
                    {\lower.5ex\hbox{$\sim$}}}}
\def\gapp{\mathrel{\rlap{\raise.5ex\hbox{$>$}}
                    {\lower.5ex\hbox{$\sim$}}}}    

\section{Introduction}
The observation of neutrino oscillation stands for one of the windows facing towards new physics beyond the Standard Model~(SM). In fact it is essentially the one found in the laboratory that has been established beyond doubt, thanks to the neutrino oscillation experiments \cite{Fukuda:2001nk,Ahmad:2002jz,Ahmad:2002ka,Eguchi:2002dm,Abe:2008aa,Abe:2011sj,Abe:2011fz,An:2012eh,Ahn:2012nd}. In order to explain this oscillation phenomenon and it's associated consequences it is imperative to include the right handed counter part of the SM neutrino to the SM fermion sector, which leads to make the SM neutrinos massive through the type-I seesaw mechanism. One appealing feature of the classical seesaw mechanism is that it can explain the observed baryon asymmetry of the Universe~(BAU) via the mechanism of leptogenesis, as pointed out by Fukugita and Yanagida \cite{Fukugita:1986hr}. The predominance of matter over antimatter has been evidenced by many experimental observations \cite{Hinshaw:2012aka,Aghanim:2018eyx}. This excess of matter in the present Universe is quantified by a quantity called baryon to photon ratio ($\eta_B$), which has been reported by the recent Planck satellite experiments as \cite{Aghanim:2018eyx,ade2016planck} 
\begin{equation}
\eta_{B}=\frac{n_{B}-n_{\overline{B}}}{n_{\gamma}}= (6.02 - 6.18)\times 10^{-10}.
\end{equation}  
where, $n_B$, $n_{\bar B}$ and $n_\gamma$ are respectively the number densities of baryons, anti-baryons and photons.

Leptogenesis is a mechanism by which some lepton-number-violating processes \cite{Minkowski:1977sc,Mohapatra:1979ia,Ellis:1992nq,Rubakov:1996vz} produce a lepton asymmetry which is subsequently converted into a baryon asymmetry through the non-perturbative $B+L$ violating but $B - L$ conserving sphaleron processes of the SM~\cite{PhysRevD.49.6394,DOnofrio:2012phz}. Based on the type-I seesaw~\cite{Mohapatra:1979ia,Magg:1980ut,Schechter:1980gr,Ma:1998dn} there has been proposed numerous frameworks (see for instance Ref. \cite{Barbieri:1999ma,Buchmuller:1999cu,Buchmuller:2003jr,Nardi:2006fx,Rahat:2020mio,Dev:2017trv,Blanchet:2006be,Mukherjee:2018fms,Narendra:2020hoz}) which are motivated by the studies of neutrino mass and a possible explanation for the BAU through the process of leptogenesis. Based on the temperature regime at which leptogenesis is supposed to take place, there exist several field theoretically consistent frameworks which in particular deal with the characteristic features of lepton asymmetry generation at a certain temperature \cite{Buchmuller:2004nz,Giudice:2003jh,Davidson:2008bu,Plumacher:1996kc,Buchmuller:2004tu}. Thermal leptogenesis is motivated by the scenario of lepton asymmetry production at a comparatively higher temperature which is in agreement with the type-I seesaw scale ($\mathcal{O}(10^{15})$). However, such a scenario predicts the RHN mass to be close to the GUT scale, which is difficult to probe at the LHC. In this context the low scale seesaw models can in principle explain the BAU through leptogenesis by the decay of a low scale (typically TeV scale) RHN. At such low scale the lepton asymmetry receives a resonant enhancement, a scenario termed as resonant leptogenesis which can be found in Refs. \cite{Pilaftsis:2003gt,Deppisch:2010fr}. 

 As the lepton asymmetry parameter solely depends on the Dirac Yukawa coupling matrix, it is essential to construct the same containing the low energy observables explaining neutrino mass and mixing. Such Yukawa coupling matrix can be obtained with the help of a well known formalism known as Casas-Ibarra~(CI) parameterisation \cite{Casas:2001sr}. There are various parameterisations available in the literature \cite{Casas:2001sr,Pascoli:2003rq}. Among them the most widely used parameterisation is the one mentioned in \cite{Casas:2001sr}. One aesthetic feature of such parameterisation is that, it can bridge the low energy inputs with high energy observations such as the estimation of the BAU. The Casas-Ibarra parameterisation introduces an orthogonal matrix (generally denoted as $R$) which in general play a key role in providing the high energy CP violation, while computing the CP asymmetry through leptogenesis. In principle this matrix is chosen to be complex by nature, considering the rotation angles to have both real and imaginary parts\footnote{Although some early literature assumed this matrix to be real, for the purpose of making the connection of low and high energy CP violation strong, see e.g., Ref. \cite{Berger:1999bg,Joshipura:1999is,Falcone:2000ib,Joshipura:2001ui,Endoh:2000hc,Rebelo:2002wj,Branco:2002kt,Ellis:2002xg,Frampton:2002qc,Endoh:2002wm,Rodejohann:2002hx,Davidson:2002em,Pascoli:2003uh,Molinaro:2008rg,Moffat:2018smo,Li:2021tlv}.}. The nature of this $R$ matrix can in principle play a decisive role in the context of leptogenesis and LFV as extensively studied in Ref. \cite{Pascoli:2003uh,Pascoli:2006ci}. Apart the influence of the orthogonal matrix on leptogenesis or LFV related calculations, there exist one more fascinating direction which is, it provides the connection between the low energy and high energy CP violation. In this regard there has been plenty of studies where such connection is discussed for different regimes of leptogenesis which solely depends on the RHN mass scale. 
 
 In this work we attempt to reappraise the aforementioned connection from a different perspective, considering a special construction of the orthogonal matrix $R$, as can be found in the literature \cite{Pascoli:2003rq,Pascoli:2006ci}, which makes the CI parameterisation as a whole very predictive. We study the deterministic nature of these two $R$ matrices, in the context of having a viable Leptogenesis parameter space. According to this special construction, the $R$ matrix can further be expressed as an exponential of a skew-symmetric matrix, the elements of which needs a deeper evaluation\footnote{Impotance of such evaluation has also been addressed in a very recent work \cite{Arias-Aragon:2022ats}.}.  
This article primarily aims to emphasize the role of the low energy CP phases in high energy CP violation, assuming the Dirac CP phase being the only source of CP violation in the low energy sector.  A detailed study in this regard have become essential after the announcement made by T2K \cite{Abe:2019vii} in the recent year. We try to highlight the importance of the Dirac CP phase for the given choices for the orthogonal matrix, making the CI parameterisation different from the usual one\footnote{In the usual CI parameterisation the orthogonal matrix $R$ is parameterised as a complex orthogonal matrix, as can be found in \cite{Casas:2001sr}.}. The characteristic features of these two different $R$ matrices lead to interesting phenomenologies which can be verified in the low energy experiments. Especially the predictions on the low energy phase can be probed in the neutrino factories. It is worth mentioning here that, flavored leptogenesis can lead to baryogensis for any value of the PMNS phases (Dirac or Majorana) as claimed in \cite{Davidson:2007va}. Thus, it is not definite that the baryon asymmetry obtained through flavor efffects in leptogenesis would restrict the values of the PMNS phases. Most importantly, the existence of such restriction is also decided by certain parametrization as stated in \cite{Davidson:2004wi}, which is the matter of investigation here. The constraint on a parameter which we would call as the skew symmetric matrix element (SSME)\footnote{The elements ($a,b,c$) of this skew symmetric matrix {\bf A} were also named as leptogenesis  CP-violation (CPV) parameter as can be found in the Ref.~\cite{Petcov:2006pc}} which is explicitly involved in the CI parametrisation can be probed through the different LFV experiments. 

We have chosen the canonical type-I seesaw mechanism having three RHNs. The scale of the leptogenesis is decided by the mass scale of the RHNs, the decay of which is supposed to source the lepton asymmetry.  Taking into account the leptogenesis constraints on the relevant parameters we show the predictions on low energy parameters which may have relevance in the upcoming neutrino oscillation experiments. In addition  we explore here the role of the Dirac CP phase in bringing a non-zero lepton asymmetry considering the RHN mass scale to be around $10^{8}$ GeV. The observational bound on $\eta_B$ introduces a limit on the range of the Dirac CP phase, the lightest neutrino mass and the SSME. We consider a nearly degenerate spectrum for the RHN masses to account for the case of leptogenesis. The required amount of mass degeneracy is found to differ by an order of magnitude for each cases of the orthogonal matrix considered here. 

This article is organized as follows. In Section~\ref{CI} we discuss about different kinds of CI parameterisations we are going to explore. Here we also provide a brief discussion on the neutrino Yukawa Lagrangian and the neutrino mass and mixing parameters. Section~\ref{baryolepto} is provided with the required prescriptions for the calculation of lepton asymmetry. Section~\ref{pheno} is kept for the detailed phenomenology as obtained owing to the choices of these two parameterisations. In~Section~\ref{sec:beq} we discuss the evolution of RHN population and lepton asymmetry with the help of Boltzmann Equations. In Section~\ref{lfv} we discuss the required parameter space which are tightly constrained in view of LFV data.  Finally we highlight the conclusions of our analysis in Section~\ref{conclusion2}. In Appendix \ref{appendix2} we have reported a general conclusion on the viability of leptogenesis under these considerations of the rotational matrix structure for a wide range of RHN mass, the SSME and the lightest neutrino mass.

\section{Type-I seesaw and Casas-Ibarra parameterisation}\label{CI}
Extension of the SM fermion sector by a pair of RHNs (the minimalistic scenario \cite{Ibarra:2003up}) are no more a choice but has become essential to explain the tiny neutrino mass through the type-I seesaw mechanism. For such a minimalistic scenario with two RHNs one should obtain the lightest neutrino mass to be vanishing. However we chose the option of having three RHNs where we get the lightest neutrino mass eigen value to be non-zero. In the type-I seesaw mechanism the heavy RHN couples with the SM lepton and the Higgs doublet through Yukawa like interaction. The coupling governing such interaction serves as the key role in offering the tiny neutrino masses through the canonical type-I seesaw mechanism. The same Yukawa coupling also governs the necessary interactions violating CP symmetry and thereby potentially explaining the origin of matter-antimatter asymmetry via the process of leptogenesis \cite{Fukugita:1986hr}. To obtain the neutrino Yukawa coupling from a model independent perspective Casas-Ibarra parameterisations have been playing promising role. For such example related to this please refer to \cite{Xing:2009vb,Chakraborty:2020gqc,Xing:2020erm,Borah:2020wyc}.  

In this section we briefly describe the type-I seesaw mechanism having three RHNs, a framework where all the three active neutrinos get tiny but non-zero masses.  We also detail the lepton mixing matrix and the low energy neutrino observables subsequently.           
\subsection{Type-I seesaw and neutrino mass}\label{type_I}
The Yukawa Lagrangian generating the light neutrino mass through the type-I seesaw can be cast into,
\begin{equation}
-\mathcal{L} = Y_{\nu}^{ \ell i} \,\overline{L_\ell} \,\widetilde{H} \,N_{R_i} + M_{R} \,\overline{(N_{R_i})^{c}} \,N_{R_i} + h.c.
\end{equation}
with, $\ell,i$ being respectively the flavor and generation indices for three generations of leptons and RHNs. The above Lagrangian generates a Majorana mass matrix for the left-handed neutrinos of the form:
\begin{equation}
m_\nu = -\left(Y_{\nu} v \right) M_{\text{diag}}^{-1} \left(Y_\nu v\right)^{\rm T},
\label{nondiaglightmass}
\end{equation}
where, $M_{\text{diag}} \equiv \text{diag}(M_{1},M_{2},M_{3})$ and $v$ being the SM Higgs VEV. The light Majorana neutrino mass matrix $m_{\nu}$ can be diagonalised approximately by the unitary PMNS matrix $U_{\text{PMNS}}\simeq U$, as follows
\begin{equation}
U^{\dagger} \,m_{\nu} \,U^{*}=m_{\text{diag}}~,
\label{diaglightmass}
\end{equation}
where, $m_{\text{diag}} \equiv \text{diag}(m_{1},m_{2},m_{3})$, with the mixing matrix $U$ having the following form, 
\begin{equation}
U=\left(\begin{array}{ccc}
c_{12}c_{13}& s_{12}c_{13}& s_{13}e^{-i\delta}\\
-s_{12}c_{23}-c_{12}s_{23}s_{13}e^{i\delta}& c_{12}c_{23}-s_{12}s_{23}s_{13}e^{i\delta} & s_{23}c_{13} \\
s_{12}s_{23}-c_{12}c_{23}s_{13}e^{i\delta} & -c_{12}s_{23}-s_{12}c_{23}s_{13}e^{i\delta}& c_{23}c_{13}
\end{array}\right) U_{\text{M}}
\label{PMNS}
\end{equation}
where, we define $c_{ij} = \cos{\theta_{ij}}, \; s_{ij} = \sin{\theta_{ij}}$ as the $\sin$ and $\cos$ of the three mixing angles for three lepton generations and $\delta$ as the leptonic Dirac CP phase. The diagonal matrix $U_{\text{M}}=\text{diag}(1, e^{i\alpha}, e^{i \beta})$ contains the undetermined Majorana CP phases $\alpha, \beta$. One can express the neutrino mass eigenvalues as $m^{\text{diag}}_{\nu} 
= \text{diag}(m_1, \sqrt{m^2_1+\Delta m_{21}^2},\sqrt{m_1^2+\Delta m_{31}^2})$ for normal hierarchy (NH) and $m^{\text{diag}}_{\nu} = \text{diag}(\sqrt{m_3^2+\Delta m_{23}^2-\Delta m_{21}^2}$, $\sqrt{m_3^2+\Delta m_{23}^2}, m_3)$ for inverted hierarchy (IH). A clear idea on these mass and mixing observables can be found from the global fit oscillation parameters as presented in Table \ref{tabdata}.
\begin{table}
\begin{center}
\begin{tabular}{|c|c|c|c|c|}
\hline
Parameters & Normal hierarchy & Best fit (NH)& Inverted hierarchy & Best fit (IH)  \\
\hline \hline
$\sin^2\theta_{23}$ &   0.415 - 0.616 & 0.573   & 0.419 - 0.617& 0.575 \\
\hline
$\sin^2\theta_{12}$ &   0.269 - 0.343 & 0.304   &  0.269 - 0.343 & 0.304  \\
\hline
 $\sin^2\theta_{13}$ & 0.02032 - 0.02410&  0.02219  & 0.02052 - 0.02428 & 0.02238   \\
\hline
$\Delta m^2_{21}/ 10^{-5}$ eV$^2 $ & (6.82 - 8.04)& 7.42  &   (6.82 - 8.04)& 7.42\\
\hline
$\Delta m^2_{31} /10^{-3}$ eV$^2$ & (2.435 - 2.598) & 2.498    & - (2.581 - 2.414) & 2.517  \\
\hline
$\delta_{CP}/ ^{\circ}$ &120 - 369&  197  & 193 - 352 & 282   \\
\hline
\end{tabular}
\label{tabdata}
\caption{Standard inputs used in the analysis, taken from \cite{Esteban:2020cvm}.}
\end{center}
\end{table}


Using the CI parameterisation $Y_\nu$ can be expressed as the product of a unitary matrix $U_L$ and some other matrices where $U_L$ can be identified with the PMNS matrix $U$. It is now evident that CI formalism bridges the Dirac Yukawa coupling with the low energy parameters, present in the lepton mixing matrix. Hence, one can expect the low energy neutrino parameters to get restriction from the baryon asymmetry constraint. 
From Eqs.\,\ref{nondiaglightmass} and \ref{diaglightmass}, the Yukawa coupling matrix can be expressed as,
\begin{equation}
Y_{\nu} = \frac{i}{v} \,U \,m_{\text{diag}}^{1/2} \,R \,M_{\text{diag}}^{1/2}\,,
\end{equation}
where, in general $R$ is a complex orthogonal matrix, which implies $R R^{T}={\bf \mathbb{I}}$. Here we consider two different forms of orthogonal matrix: i) $R= {\bf O} \,e^{i {\bf A}}$ and ii) $R={\bf O} \,e^{{\bf A}}$, where ${\bf A}$ is a skew symmetric matrix and {\bf O} is in principle an orthogonal matrix. However, for simplicity here we choose {\bf O} to be an identity matrix\footnote{Relaxing this assumption makes the scenario less predictive, in the context of investigating the role of Dirac CP violation in baryogenesis. The baryon asymmetry criteria does not impose any restriction on the range of Dirac CP phase.}. This choice is also motivated by the near quasi degeneracy (QD) among the light neutrino masses, that we consider in this work - quasi degeneracy in normal hierarchy. As the effect of {\bf O} can be absorbed in the PMNS matrix and $U$ and $U.{\bf O}$ would lead to the same physics (see Sec. 2 of Ref. \cite{Pascoli:2003rq}), one can work with {\bf O} $={\bf \mathbb{I}}$. This kind of assumption is driven by the presence of {\bf O}$(3)$ symmetry in the lepton sector \cite{Branco:1998bw}. Simultaneously, it is spontaneous to consider a quasi-degenerate spectrum for the heavy Majorana neutrinos also, $M_1 \approx M_2 \approx M_3 = M$.  Relaxing which may require, an unnatural fine-tuning between $Y_{\nu}$ and $M$ to get a nearly QD spectrum for the light neutrinos. It is also to be noted that, we perform this analysis considering near quasi-degeneracy in the normal hierarchy of neutrino mass spectrum. Both of these choices for the $R$ matrix especially give rise to exponential growth in the Yukawa coupling parameter space which eventually alters the parameter space for leptogenesis when compared to the case where $R$ is simply a complex orthogonal matrix. 

In Ref.\,\cite{Pascoli:2003rq} we get the notion that the elements of {\bf A} can be related with the RHN mass~($M_{R}$) scale so as to have an appropriately ordered Yukawa coupling. To realize which one can analytically write $|Y^{ij}_\nu| \leq \sqrt{4\pi}$, which in turn imposes constraints on the scale of the RHN mass and also on the {\bf A} matrix elements for a particular seesaw scenario. Eq.\,[18] of \cite{Pascoli:2003rq} tells us that one can choose larger values of these matrix elements (here, {\bf a}), which in principle can correspond to a relatively smaller domain for the RHN mass as compared to the one associated with the classical type I seesaw scale ($10^{15}$~GeV). For a type-I seesaw scenario, there has been reported some numerical values corresponding to these SSMEs (see {\it e.g.,} \cite{Pascoli:2003rq}) in view of high scale leptogenesis for the RHN mass to be around $10^{11}$~GeV. This range of values however can further be constrained by considering different low energy experiments like LFV, the discussion of which is provided in section \ref{lfv}. It is worth noting that based on the choices of the rotational matrix mentioned above the allowed ranges of the SSMEs are different. In the following subsections we present the details of the CI parameterisations under the choices made for the orthogonal matrix $R$.


\subsection{Case-I: when $R=e^{i{\bf A}}$}
With the above choice of $R$ matrix the Yukawa coupling matrix can be written as, 
\begin{equation}\label{CI_1}
Y_{\nu} = \frac{i}{v} \,U \,m_{\text{diag}}^{1/2} \,e^{i{\bf A}} \,M_{\text{diag}}^{1/2}.
\end{equation}
In case of this complex parameterisation the orthogonal matrix $R=e^{i {\bf A}}$ can be expanded as~\cite{Pascoli:2003rq,Petcov:2005yh,Petcov:2006pc}:
\begin{equation}
e^{i {\bf A}}=1-\frac{\cosh r-1}{r^{2}} {\bf A}^{2}+i\frac{\sinh r}{r} {\bf A},
\label{eiA_expr}
\end{equation}
where, ${\bf A}$ is a real skew symmetric matrix having satisfied the feature of the $R$ matrix to be orthogonal ($R R^T = \mathbb{I}$). 
\begin{equation}
\begin{pmatrix}\label{skew}
0 && a && b \\
-a && 0 && c \\
-b && -c && 0 \\
\end{pmatrix},
\end{equation}           
with, $r=\sqrt{a^{2}+b^{2}+c^{2}}$. For reference, we name this case as the complex case and the later (Case-II) as the real case. 
 \subsection{Case-II: when $R=e^{{\bf A}}$ }
 The Yukawa matrix has the following structure with the choice $R=e^{{\bf A}}$
 \begin{equation}\label{CI_2}
Y_{\nu} = \frac{i}{v} \,U \,m_{\text{diag}}^{1/2} \,e^{{\bf A}} \,M_{\text{diag}}^{1/2}.
\end{equation}
In case of this parameterisation the orthogonal matrix $R=e^{{\bf A}}$ can be expanded in the following form \cite{Pascoli:2006ci}:
\begin{equation}
e^{{\bf A}}=1+\frac{1-\cos r}{r^{2}} {\bf A}^{2} + \frac{\sin r}{r} {\bf A},
\label{eA}
\end{equation}
where, the matrix {\bf A} and $r$ have the similar definitions as in the former case described above. 
 For simplicity and we consider here the elements of the matrix ${\bf A}$ to obey, $a=b=c= {\bf a}$, and hence one can have $r=\sqrt{3}\, {\bf a}$. Such an assumption allows for a minimal number of free parameters. For convenience, in the rest of this article we name this parameter {\bf a} to be SSME as we mention in the introduction. This kind of equality among these three elements can be found in a very recent work \cite{Konar:2020vuu}. Derivation of Eqs.\,\ref{eiA_expr} and \,\ref{eA} can be found in the Appendix\,\,\ref{appendix_complex}.  
\section{Leptogenesis}\label{baryolepto}
In a temperature regime where all the lepton flavors are distinguishable, it is instructive to consider the flavor dependent as well as resonant leptogenesis approach. Although for a pure resonant scenario (see e.g. \cite{Pilaftsis:2003gt}), one has to work with a highly degenerate RHN mass spectrum which is not the case here\footnote{For resonant leptogenesis to work, the mass difference between the RHN states have to be of the order of their decay width. This requirement is not satisfied with the chosen RHN mass spectrum here, as $1 - x_{ij} \neq \frac{1}{8\pi} Y_\nu^\dagger Y_\nu $. The equality between these two parametric expressions is an essential condition for resonant leptogenesis to take place. Authors in \cite{Dev:2015wpa} have shown that, if the required mass difference is driven by renormalization group (RG) running, then it is difficult to achieve successful leptogenesis within a minimal radiative scenario. In such case one requires non-minimal extension to rescue leptogenesis parameter space. It is to note that the present set-up is free from such scenario.}. The inclusion of flavor effects on leptogenesis provides a very essential modification for the calculation of the final baryon asymmetry (please see Ref. \cite{Blanchet:2006be,Abada:2006ea,Nardi:2006fx,Dev:2017trv}), as compared with the calculation in the unflavored scenario. In such scenario the CP asymmetry generated from the decay of the lightest RHN can be expressed as \cite{Adhikary:2014qba, Dolan:2018qpy}\footnote{This prescription for lepton asymmetry do not include the effect of thermal masses of final state particles, which becomes slightly influential only at higher temperature ($T>> M$). However, the influence is negligible in the context of the order of magnitude of the lepton asymmetry ($\epsilon$) as shown in \cite{Giudice:2003jh}.},
\begin{align}\label{Eq:asymmetry}
    \epsilon_{i}^{\ell} = & \frac{1}{8\pi \left(Y_\nu^{\dagger} Y_\nu\right)_{ii}}\sum_{j \neq i} \text{Im}\left[\left(Y_{\nu}^\dagger Y_{\nu}\right)_{ij}\left(Y_\nu^{\dagger}\right)_{i\ell}\left(Y_{\nu}\right)_{\ell j}\right]  \left[ f(x_{ij}) + \frac{\sqrt{x_{ij}} \left(1 - x_{ij}\right)}{\left(1- x_{ij}\right)^2 + \frac{1}{64 \pi^2} \left(Y_{\nu}^\dagger Y_{\nu}\right)_{jj}^2}\right]  \nonumber\\ 
    & + \frac{1}{8\pi \left(Y_\nu^\dagger Y_\nu\right)_{ii}}\sum_{j \neq i} \frac{ (1- x_{ij}) \text{Im}\left[\left(Y_{\nu}^\dagger Y_{\nu}\right)_{ij}\left(Y_\nu^{\dagger}\right)_{i\ell}\left(Y_{\nu}\right)_{\ell j}\right] }{\left(1- x_{ij}\right)^2 + \frac{1}{64 \pi^2} \left(Y_{\nu}^\dagger Y_{\nu}\right)_{jj}^2} + \mathcal{O} \left(Y_{\nu}^6\right),
   \end{align}
with the following definition for the loop function 
$ f(x_{ij}) = \sqrt{x_{ij}} \left[1 - (1+x_{ij})\text{ln}\left(\frac{1-x_{ij}}{x_{ij}}\right) \right]$ 
where, $x_{ij} = \left(\frac{M_{j}}{M_{i}}\right)^2$. One can define $Y_{\nu}$ as the complex Dirac Yukawa coupling, derived in a basis where the RHNs are in the diagonal mass basis.

With the above prescription for lepton asymmetry one can write the analytically approximated solution (from the set of Boltzmann equations given by Eq.~\ref{eq:beq}) for the baryon to photon ratio \cite{Pilaftsis:2003gt,Deppisch:2010fr,Bambhaniya:2016rbb} as,
 \begin{equation}\label{Eq:bau}
 \eta_B \simeq -3\times 10^{-2} \sum_{\ell,i}\frac{\epsilon_{i \ell}}{K_\ell^{\text{eff}}\text{min}\left[z_c,1.25 \, \text{Log}\,(25 K_\ell^{\text{eff}})\right]} ~,
\end{equation}  
where $z_c = \frac{M_i}{T_c}$ and $T_c \sim 149$ GeV, \cite{Bambhaniya:2016rbb} is the critical temperature, below which the sphalerons freeze out \cite{PhysRevD.49.6394,DOnofrio:2012phz}. Here,  $K_\ell^{\text{eff}} = \kappa_\ell \sum_{i} K_i B_{i\ell}$, with $K_i = \Gamma_i / \zeta(3)H$, denoting the washout factor and $\Gamma_i = \frac{M_i}{8\pi}(Y_{\nu} Y_{\nu}^\dagger)_{ii}$ as the tree level heavy-neutrino decay width. The Hubble rate of expansion at temperature $T\sim M_i$ can be expressed as ,
\begin{equation*}
H = 1.66 \sqrt{g^*}\frac{M_i^2}{M_{\text{Pl}}}\;\; \text{with}\;\;\; g^* \simeq 106.75 \;\;\;\text{and}\;\;\; M_{\text{Pl}} = 1.29 \times 10^{19} \,\text{GeV}.
\end{equation*}
 Here, $B_{i\ell}$'s are the branching ratios of the $N_i$ decay to leptons of $\ell^{th}$ flavor : $B_{i\ell} = \frac{|Y_{\nu_{i\ell}}|^2}{(Y_{\nu}Y_{\nu}^{\dagger})_{ii}}$.
Including the Real Intermediate State~(RIS) subtracted collision terms one can write the factor $\kappa$ as,
\begin{equation}
\kappa_\ell = 2 \sum_{i,j j \neq i} \frac{\text{Re}\left[(Y_{\nu})_{i\ell}(Y_{\nu})_{ j\ell}^* \left(Y_{\nu} Y_{\nu}^\dagger\right)_{ij}\right]+ \text{Im}\left[\left(\left(Y_{\nu}\right)_{ i\ell} (Y_{\nu})_{ j\ell}^*\right)^2\right]}{\text{Re}[(Y_{\nu}^\dagger Y_{\nu})_{\ell \ell} \{(Y_{\nu} Y_{\nu}^\dagger)_{ii} + \left(Y_{\nu} Y_{\nu}^\dagger\right)_{jj}\}]}\times\left(1-2i \frac{M_i-M_j}{\Gamma_i + \Gamma_j}\right)^{-1}.
\end{equation} 
Once we numerically evaluate $Y_\nu$, they can be further used in the above equations for computing the baryon to photon ratio. In the following section we are going to present the methodology involved for pursuing the parameter space extraction for leptogenesis using the two different CI formalisms discussed in the previous section \ref{CI}. 
\section{Numerical analysis}\label{pheno}
As mentioned earlier one of the aims of choosing these two forms of the $R$ matrix is to investigate the allowed ranges of the skew symmetric matrix element ({\bf a}) in view of leptogenesis in the type-I seesaw. On the other hand, the range of the SSME which is allowed by a successful leptogenesis scenario can be indirectly verified in a low energy experiment such as lepton flavor violation (see for instance\cite{Pascoli:2006ci}). Note that this range can be very specific for each variety of the $R$ matrix considered in this analysis. A nearly degenerate spectrum for the RHN masses is required for yielding a non-zero lepton asymmetry. Accordingly, for both the cases we take a nearly degenerate spectra for the first two RHN mass states, $M_1 \approx M_2$, keeping the third one ($M_3$) relatively more massive. This assumption leads us to write $ M_1, M_2, M_3 = 10^8 , 10^8 + \Delta M, 10^9 $ GeV\footnote{How a different regime of the RHN mass scale impacts this analysis, is provided in the appendix \ref{appendix2}.}. We choose a range of this mass splitting $\Delta M$, for instance, to be varied from ($0.001 - 0.1$)~GeV through out this analysis. This small mass splitting among the RHN mass states is not fine tuned and can be a natural choice in particular when these special kinds of $R$ matrices are considered. For simplicity we have used the best fit central values for the low energy neutrino parameters namely, the mixing angles $\theta_{ij}$ and the mass squared splittings $\Delta m_{ij}^2$, while varying the Dirac CP phase from $0 \,-\, 2\pi$\footnote{We emphasize to comment that a random scan over these neutrino mixing parameters does not bring significant modifications in the results of this analysis. That is the reason behind the choices of these parameters at their best fit central values.}. Also we restrict this analysis for normal hierarchy (NH) of neutrino masses keeping in mind the recent preference for NH~\cite{Capozzi:2018ubv}. As mentioned earlier we focus on the assumption that the source of CP-violation arises only from the low energy phase $\delta$ and hence to ensure this we choose the Majorana phases $\alpha$ and $\beta$ to be zero. As it is evident from the expressions of the $R$-matrices that the exponential rise as a function of the SSME can bring a large enhancement in the Yukawa coupling which may not be desirable. Therefore,  care has been taken to ensure the perturbativity limit of the Yukawa couplings ($Y_\nu$) which is in principle realized by writing $|(Y _\nu )_{ij}| \leq \sqrt {4 \pi}$.

 With the aforementioned structures of the orthogonal matrix and the input parameters we proceed for the numerical computation for the lepton and baryon asymmetry. While doing so we investigate the possible constraints which might have been imposed by the baryon asymmetry on the parameters namely, the SSME ({\bf a}), the lightest neutrino mass ($m_1$), and the Dirac CP phase ($\delta$). We present these constraints in the following manner i) the predictions on low energy parameters namely, the lightest neutrino mass in case the neutrino masses obey normal hierarchy and the Dirac CP phase, ii) the connection between the low energy phase $\delta$ and the baryon asymmetry, and iii) the role of the Dirac CP phase in individual flavored asymmetries,  considering the present forms of the rotational matrix $R$. We provide the related discussions on the relevant phenomenologies for each individual cases of the $R$ matrix concerned in two different subsections. 

 In the following subsections we present the notable features of these two parameterisations in the context of baryogenesis through leptogenesis.  As mentioned earlier, we present this analysis in two different contexts, one focusing on the low energy predictions in the light of these special forms of $R$ matrix and another, finding their implications in the theoretical study related to the lepton flavor violating decay $\mu \rightarrow e \gamma$. The appearance of two different kinds of the rotational matrix ($R$) imposes constraints on the Dirac CP phase very differently. Interestingly the chosen ranges for the lightest neutrino mass, the Dirac CP phase and the SSME appear to be restricted by the requirement of satisfying the baryon to photon ratio constraint. 
\subsection{Predictions on parameters: when $R=e^{{\bf iA}}$}
Among the input parameters $m_{1} ~\text{and}~ \delta$ can in principle be assumed as low energy parameters as they can be probed in the low energy experiments. Whereas the remaining two ${\bf a}~\text{and}~\Delta M$ can be considered as high energy parameters as they are supposed to be involved in the neutrino mass generation mechanism at high energies through the construction of the Yukawa matrix (Eqs. \ref{CI_1} and \ref{CI_2}). We examine the high energy parameters (${\bf a}~\text{and}~\Delta M$) by fixing a set of different values of low energy parameters namely the lightest neutrino mass and the Dirac CP phase and vice-versa. As mentioned earlier we focus on the heavy RHN mass regime to be of the $\mathcal{O}(10^{8})$~GeV, at such temperature scale the three lepton flavors ($e, \mu \; \text{and}\;\tau$) are completely distinguishable, implying a fully flavored regime for leptogenesis.           

\begin{figure*}
\begin{center}
\includegraphics[scale=0.27]{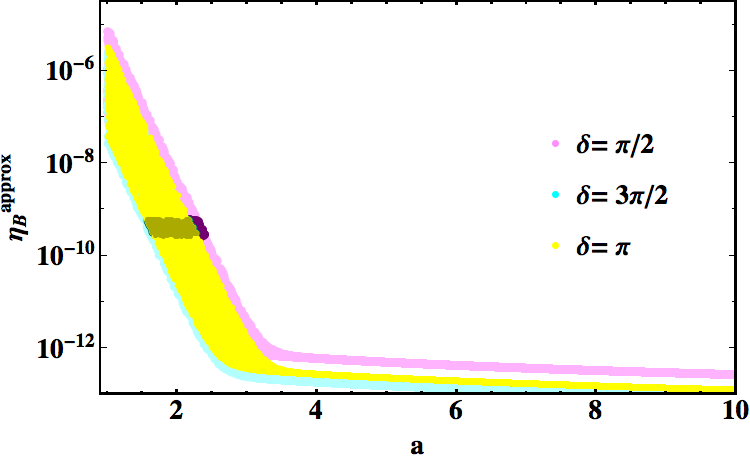}
\includegraphics[scale=0.27]{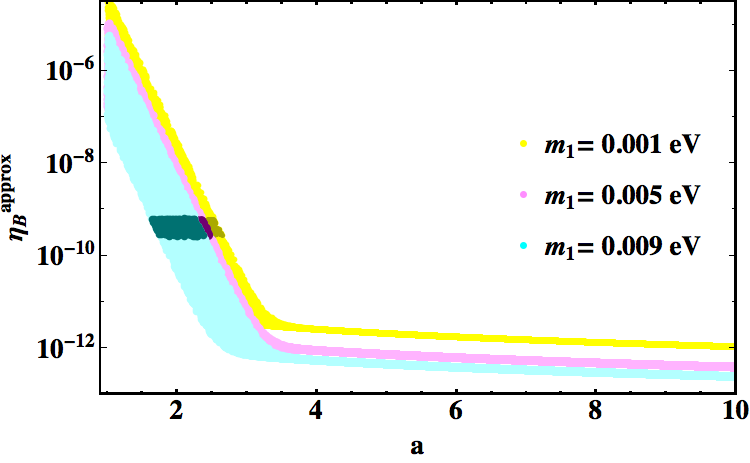}    
\caption{In the left (right) panel we show the variation of $\eta_B$ as a function of the parameter \textbf{a} fixing the Dirac CP phase $\delta$ to be at different benchmark values and the $m_{1}$ respectively. We fix $m_1 = 0.009$~eV and $\delta = \pi/2$ for the left and right panel respectively. The darkened regions exhibit the constrained range for \textbf{a}. The relevant explanation is provided in the text.}
\label{eiA_high}
\end{center}
\end{figure*}

In Fig.~\ref{eiA_high} we fix $m_1$, and $\delta$ at different set of values and varied the parameter {\bf a} from $1-10$ along with the mass splitting $\Delta M$ from ($0.001-0.1$)~GeV. 
We calculate the lepton asymmetry for some benchmark values of the Dirac CP phase $\delta=3\pi/2, \pi, \pi/2$, and for a particular value of $m_{1}=0.001$~eV. Here we notice that for decreasing values of $\delta$ the upper bound on {\bf a} gets shifted towards larger values of {\bf a} to meet the observed $\eta_B$.  A similar kind of observation we have when we repeat this analysis for some benchmark values of the lightest neutrino mass. In the right panel of Fig.~\ref{eiA_high} we fixed $m_1$ at different values setting $\delta$ at $\pi/2$ to see the constraint on {\bf a} coming from the observed $\eta_B$.  We see that the allowed parameter space for the observed $\eta_B$ (the darkened regions in each plot) restricts the range of {\bf a} to have a narrow region. For an RHN mass regime $\mathcal{O}(10^{8})$~GeV, we report a maximum value of {\bf a} to be close to $2.42$.  From the left panel of this figure it is clear that the highest {\bf a} value one can have in this set up is for a smaller value of $\delta$, which is $\pi/2$. In the right panel one can see that, for the lightest neutrino mass $m_1$ to be $0.001$ eV, the allowed {\bf a} value is largest when compared to the {\bf a} values for the other two choices of $m_1$. Note that a particular allowed value of {\bf a} corresponds to a specific value for the heavy RHN mass as also mentioned in \,\cite{Pascoli:2003rq}. Although, we do not see such restrictions on the RHN mass splitting $\Delta M$ in order to satisfy the $\eta_B$ constraint, as long as the chosen range of $\Delta M$ is provided as mentioned before. 

\begin{figure*}
\begin{center}
\includegraphics[scale=0.27]{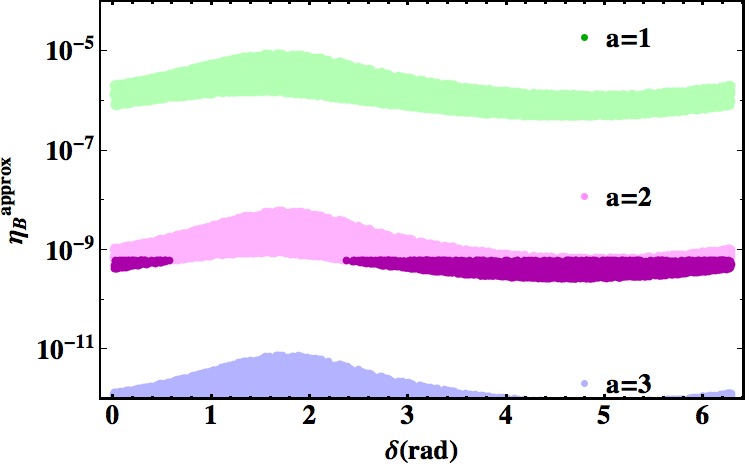}
\includegraphics[scale=0.27]{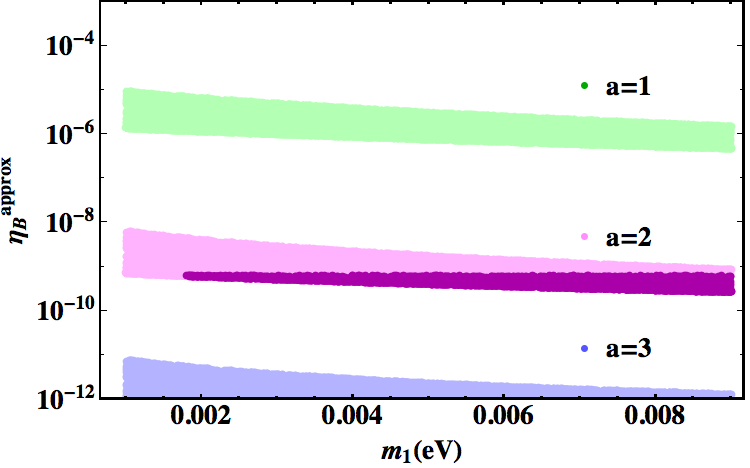}    
\caption{In the left (right) panel we show the variation of $\eta_B$ as a function of the Dirac CP phase $\delta$ ($m_{1}$ for NH) for different benchmark values of the parameter \textbf{a}. The darkened points exhibit the constrained values of Dirac CP phase and the lightest neutrino mass for the chosen hierarchy.}
\label{eiA_low}
\end{center}
\end{figure*}

\begin{figure*}[h]
\begin{center}
\includegraphics[scale=0.18]{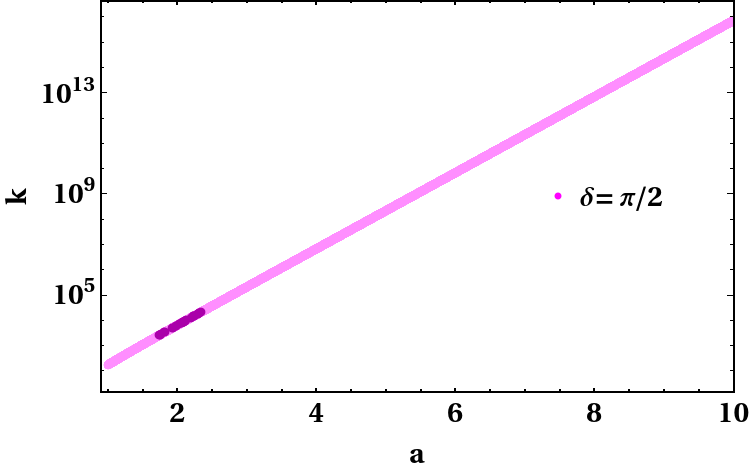}    
\includegraphics[scale=0.18]{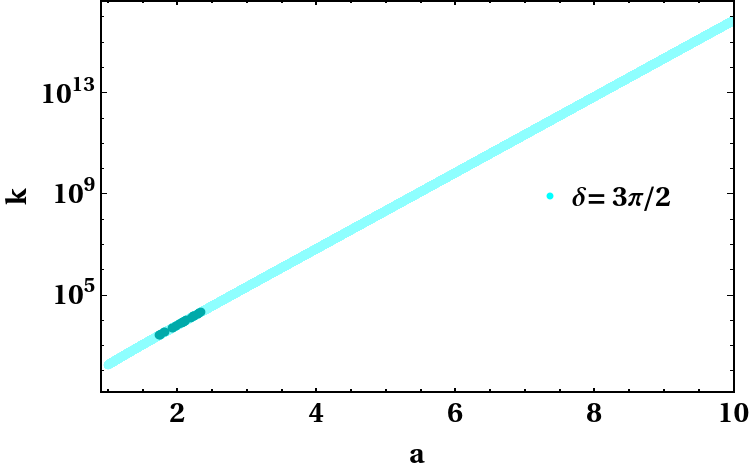}  
\includegraphics[scale=0.18]{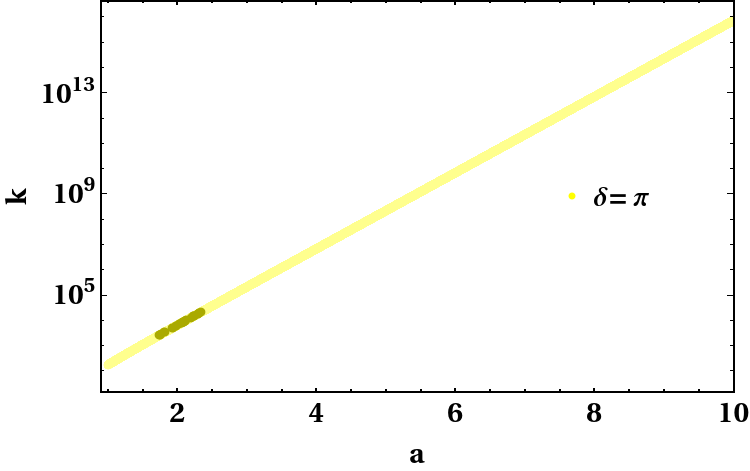}  
\caption{Shows the washout order w.r.t. the SSME {\bf a} for the complex $R$ case.  The darker regions in each figure indicates the allowed order for the washout to satisfy the observed baryon to photon ratio.}
\label{fig:washout_complex}
\end{center}
\end{figure*}
From Fig.~\ref{eiA_low} we observe that selecting some benchmark values of {\bf a} from this range {\bf ( 1\, -\, 10)} restricts the Dirac CP phase and the range of the lightest neutrino mass to achieve the $\eta_B$ constraint.
In this figure we fix {\bf a} and $\Delta M$ and vary the low energy parameters $\delta$ and $m_{1}$ from (0-2$\pi$) and (0.001-0.009)~eV respectively. The benchmark values for {\bf a} are set to be {\bf a}=1, 2, 3 keeping $\Delta M = 0.001$~GeV. It is evident from the left side of this figure that, for $\delta = \pi/2$ one obtains a maximum $\eta_B$ for any choice of {\bf a}. We also see that larger values of {\bf a} decreases $\eta_{B}$ from it's observed value (say for {\bf a}=3). This might be due to the reason that, the larger value of {\bf a} enhances the order of Yukawa couplings (having a hyperbolic kind of dependency) which, further increases the amount of washout ($K$) well enough to suppress the order of the final baryon asymmetry.  To show a clear estimate of the washout order in Fig.~\ref{fig:washout_complex} we present the washout parameter as a function of the SSME parameter {\bf a} fixing different values of the Dirac CP phase. The order of the washout is found to be insensitive to the choice on $\delta$, but very much sensitive to {\bf a}. The main restriction on the SSME range comes from the huge washout only. In these figures the darker regions imply for the largest washout that the system should not exceed to account for the observed baryon to photon ratio. One can notice here that for {\bf a}=2, $\delta$ prefers the range from ($0 - 2\pi$) except those points which are around $\pi/2$. The right panel of Fig.~\ref{eiA_low} evinces that for {\bf a}=2,  the entire range of $m_{1} \geq 0.0018$~eV can account for the observed $\eta_B$.         

 Thus it is understood that, for ${\bf a} = 2$, one should have two ranges for $\delta$ one from $(0 - 0.8)$~rad and other from $(2.3 - 6.28)$~rad as allowed by the observed $\eta_B$. Similarly, the right panel of this figure shows the restriction on the lightest neutrino mass which has been varied from $(0.001 - 0.009)$~eV. The baryon to photon ratio sets a lower bound on the lightest neutrino mass to be around $0.0018$~eV, for ${\bf a }= 2$.


\subsection{Predictions on parameters: when $R=e^{{\bf A}}$}
\begin{figure*}
\begin{center}
\includegraphics[scale=0.40]{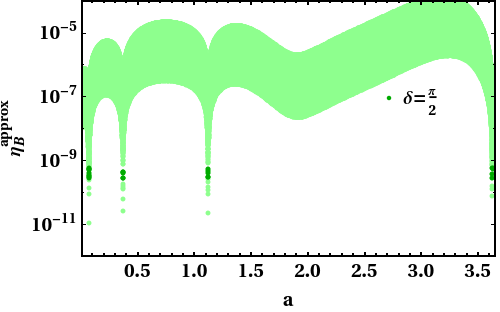}
\includegraphics[scale=0.40]{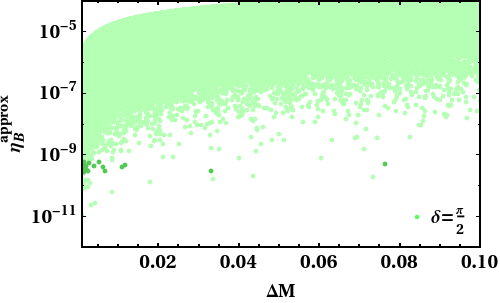}  
\caption{$\eta_B$ as a function of the parameter \textbf{a} (left) and $\Delta M$ (right) fixing $m_{1} \,= \,0.008$\,eV along with $\delta = \pi/2$. It is to note here that an odd multiple of $\pi/2$ for $\delta$ does not bring any major change in the analysis for this real $R$ case. Thus we focus here to show the variation of $\eta_B$ only for $\delta = \pi/2$. The darker green points satisfy the respective parameter spaces which are allowed by the bound on the observed $\eta_B$.}
\label{eA_high_1}
\end{center}
\end{figure*}

For numerically evaluating the baryon asymmetry while making use of the above $R$ matrix, we vary {\bf a} from $0.01 - 10$ and $\Delta M$ from $0.001 - 0.1$~GeV fixing $m_{1}$ at 0.008~eV.  We calculate  the lepton asymmetry for the aforementioned consideration of the input parameters and the form of $R$ matrix. Then we examine the relevant parameter space for $\eta_B$ considering a range of the Dirac CP phase from $\delta= (0 - 2 \pi)$.
For this real $R$ matrix, as expected from the Eq.\,\ref{eA} the behaviour of the parameterisation is sinusoidal.  As the value of {\bf a} is varied over the range from $0.01 - 10$, we see a periodic behaviour of $\eta_B$ w.r.t. {\bf a}, as evident in the left of Fig.~\ref{eA_high_1}. The periodicity of $\eta_B$ is repeated from $ {\bf a}\,=\,0.01 - 3.65$ and then from ${\bf a}\,=\,3.65 - 7.32$ and so on. We consider the range of {\bf a} to be $0.01 - 3.65$ for further analysis, as it is clear that the larger values of {\bf a} in this case do not alter the phenomenology much.           

For both the choices of $\delta=\pi/2~ \text{and}~3 \pi/2$ along with for $m_{1}=0.008$, we see a similar behaviour of $\eta_B$ when plotted with respect to {\bf a} in Fig.~\ref{eA_high_1}\footnote{Therefore we have exhibited the plots only for $\delta=\pi/2$.}. Here we observe an over estimation of $\eta_{B}$ for all {\bf a} values, except for the cases when ${\bf a}$ is equal to 0.01, 0.35, 1.12 and 3.65. These values of {\bf a} facilitate to meet the $\eta_B$ criteria for such choice of $\delta$ as mentioned in the Fig.~\ref{eA_high_1}. We also show the estimated $\eta_B$ as a function of $\Delta M$ as evinced in the right panel of Fig.~\ref{eA_high_1}.  We see that the baryon asymmetry constraint is satisfied for the entire range of $\Delta M$. However a preference for smaller $\Delta M \approx 10^{-3}$\,GeV is obtained which is presented by the darker green dense region in this figure.
\begin{figure*}
\begin{center}
\includegraphics[scale=0.40]{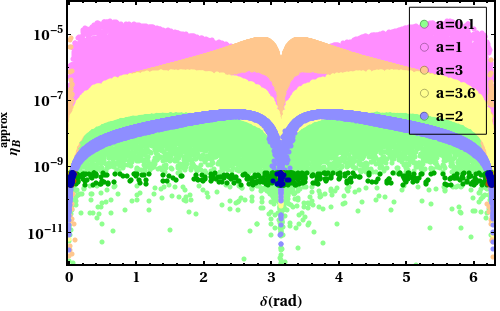}
\quad
\includegraphics[scale=0.42]{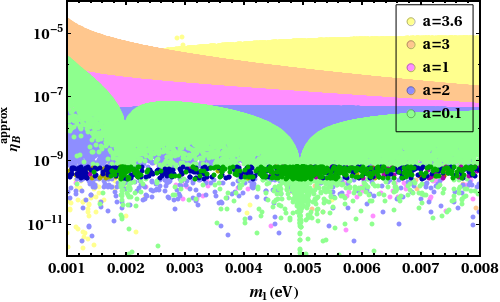}    
\caption{Variation of $\eta_B$ w.r.t. Dirac CP phase (left) and the lightest neutrino mass (right) for different benchmark values of {\bf a}. For explanation please refer to the text.}
\label{eA_delt_mL_vs_etaB}
\end{center}
\end{figure*}
To get a view on how $\delta$ influences $\et_B$ for the real $R$ case, we vary $\delta$ from $0$ to $2 \pi$ and $m_{1}$ from $0.001 \,-\, 0.008$~eV. We present the variation of $\eta_{B}$ with $\delta$ and $m_{1}$ in the left and right panels of Fig.\,\ref{eA_delt_mL_vs_etaB}, for a fixed $\Delta M$ at 0.001 GeV and considering different values of {\bf a} = 0.1, 1, 2, 3, 3.6.  In the left side of Fig.~\ref{eA_delt_mL_vs_etaB} we see that for $\delta=0, \pi$ and $2 \pi$ values the CP-asymmetry exactly vanishes\footnote{which can also be predicted from the CI parametrization as it is supposed to contain a real $R$, having no additional CP violating phases except the Dirac one.}. However, when there is slight deviation from these CP conserving values, a wider range of $\delta$ is allowed only for {\bf a = 0.1}, from the $\eta_B$ constraint. This is however not the case for the other chosen values of {\bf a}. As can be observed from the left panel of this figure, only small values of {\bf a} can prefer the entire range of $\delta$ in order to satisfy the $\eta_B$ criteria. On the other hand for all values of ${\bf a}>1$, $\delta$ is preferred to be around $0, \pi, 2 \pi$. In the right panel of Fig.\,\ref{eA_delt_mL_vs_etaB} for the same set of {\bf a} values we show the variation of $\eta_B$ with respect to the lightest neutrino mass $m_1$, which evinces that the entire range of $m_{1}$ is allowed by the observed $\eta_B$.   

\section{Boltzmann Equation for leptogenesis}\label{sec:beq}
As we discussed earlier, we consider the mass scale of RHN to be around $\mathcal{O}(10^{8})$ GeV. At this mass regime all the lepton flavors  act non identically. Hence we consider the flavor dependent Boltzmann equations (BEQ) governing the lepton charge density in individual lepton flavours.
For solving the BEQs we considered the initial conditions $\eta_\ell\left(z_{\rm in}\right)=0$ for the lepton asymmetry and $\eta _N\left(z_{\rm in}\right) = \eta_{N}^{\rm eq}$ for the RHN abundance.  A strong washout scenario is naturally accounted for in this analysis with the RHN mass regime and the order of Dirac Yukawa couplings which govern the RHN decay (inverse decay) and the $2 \leftrightarrow 2$ scatterings\footnote{It is to note that since the washout amount falls within the strong regime we have mainly focused on the inverse decay process among all other possible washout processes.}. Accordingly during the numerical solution of the equations \ref{eq:beq} we set the initial conditions. This fact can also be driven by the argument that, as the RHNs are relatively less massive, it facilitates a thermal population of these RHNs which are supposed to be produced much before the asymmetry is yielded.  The flavor-dependent BEQs relevant for leptogenesis in the present scenario can be cast into \cite{Deppisch:2010fr},

\begin{gather}\label{eq:beq}
 \frac{d\delta\eta_{N_i}}{dz} = \frac{\mathcal{K}_1(z)}{\mathcal{K}_2(z)}\left[1+ (1- K_i z)\delta\eta_{N_i}\right] ~~~ \text{with}~~~ i =1,2 \nonumber\\ 
 \frac{d\eta_{\ell}}{dz} = z^3 \mathcal{K}_1(z) K_i\left( \delta \eta_{N_i} \epsilon_{i\ell} -\frac{2}{3} B_{il}\eta_{\ell}\right), ~~~ \text {with} \;\;\;  \ell = e, \mu,\tau
\end{gather}
with $\mathcal{K}_1,~~ \mathcal{K}_2$ as the modified Bessel function of the second kind. One can define  $\delta \eta_{N} = \frac{N_{N}}{N^{\rm eq}_N} -1$ as the parameter which estimates the deviation of the RHN abundance from the Equilibrium abundance, $ K_i = \frac{\Gamma_{i}}{\zeta(3)H}$ as the washout parameter with the definition of $\Gamma_{i} \;\;\text{and}\;H$ provided in the Section~\ref{baryolepto}.  $K_i$ measures the deviation of the i'th RHN-decay rate ($\Gamma_i$) from the expansion rate of the Universe. Using the solution of the above BEQs one can estimate the baryon to photon ratio as follows:
\begin{equation}
\eta_B = -\frac{28}{51} \frac{1}{27}\sum_{\ell} \eta_\ell
\end{equation}
where, $\eta_{\ell}$ is the yield for lepton charge density obtained from the solution of the above BEQs. The factor of 28/51 arises from the fraction of lepton asymmetry reprocessed into a baryon asymmetry by the electroweak sphalerons while 1/27 is the dilution factor from photon production until the recombination epoch.
%

\begin{figure*}[h!]
\begin{center}
\includegraphics[scale=0.22]{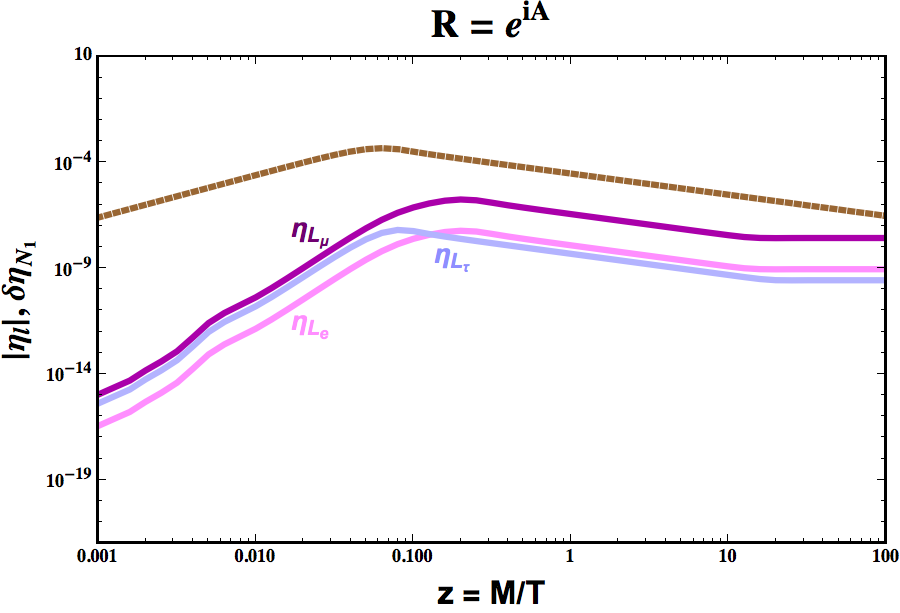}
\includegraphics[scale=0.22]{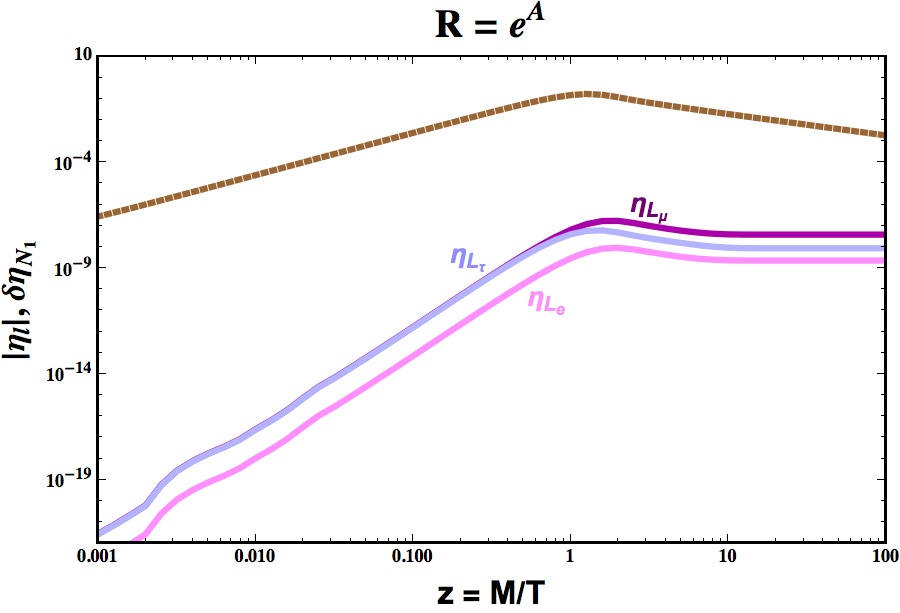}    
\caption{In the left (right) panel we show the nature of the evolution of the RHN number density and lepton asymmetry for the particular cases complex (real). The upper brown line presents the evolution of RHN population. Different colours here demonstrate the evolution of lepton asymmetry for individual lepton flavors. For this figure we set $\delta = \pi/2$ and $m_1 \,=\, 0.008$~eV. The {\bf a} values, for the two different cases involving the complex and real $R$ matrices are chosen to be {\bf a} = 2.42 and {\bf a}= 1.11 respectively.}
\label{fig:beq}
\end{center}
\end{figure*}

Fig.\,\ref{fig:beq} illustrates the deviation of RHN number density from the Equilibrium number density and the evolution of lepton asymmetry corresponding to individual lepton flavors.  Following the prescriptions of lepton asymmetry and all the washout parameters, one can notice that both explicitly depend on the Yukawa couplings~($Y_\nu$ being regulated by the $R$ matrices of interest). Therefore it is imperative to study the evolutions of lepton asymmetry resulted for the presence of different $R$ matrices.  In Fig.\,\ref{fig:beq} we present the respective evolution plots for RHN abundance and the final lepton asymmetry for the two different cases discussed above. For both of the choices of $R$ matrix we have noticed that the final asymmetry receives a major contribution from the muon flavor among the three flavors of leptons. It is almost the $95\%$ of the total asymmetry being contributed by the Yukawa couplings associated with the $\mu$-flavored leptons. The benchmark values of all the parameters chosen for making these figures have been mentioned in the figure caption. We have seen that the evolutions of the RHN number density and the flavored lepton asymmetry yield is less sensitive to the choice of the Dirac CP phase (be $\pi/2$ or $3\pi/2$) and hence we set this value at $\pi/2$ for showing these evolution plots. The reason behind the choices of {\bf a} values as mentioned in the figure caption is the following. We have chosen the maximum possible $a$ values for each individual cases representing the different forms of the $R$ matrix. In the complex $R$ case (left of Fig.~\ref{fig:beq}) we notice that at $z \,=\, 0.08$ the equilibrium abundance for the RHNs starts to fall and the maximum asymmetry for the muon and electron flavor is produced at $z\,=\,0.2$. However, the asymmetry due to all the flavors have reached a saturation at around $z\,=\,10$. The final baryon asymmetry obtained at around $z\,=\,10$ for the complex $R$ case with the aforementioned BP set of values is $3.65 \times 10^{-10}$. This evolution with the temperature regime represented by the $z$ parameter is a little different in the case involving the real $R$ (right side of Fig.~\ref{fig:beq}). One can notice a later $z$ value at which the RHN abundance starts to fall from the equilibrium abundance and the lepton asymmetry due to each individual lepton flavor reached a maximum value which is $z\,=\,1$. The final baryon asymmetry for the real $R$ case is estimated to be $2.67 \times 10^{-10}$ which receives nearly $77\%$ of the total contribution from the Yukawa interactions which involve the muon flavor.  For both the forms of $R$ matrix, we get the largest lepton asymmetry due to the muon flavor which can also be understood from table \ref{yukawa_table}. It is evident that, the overall Yukawa coupling due to muon flavor (in the second row) are larger than those associated with electron and tau flavors.

\begin{table}[h!]
\small\addtolength{\tabcolsep}{-7pt}
\begin{center}
\begin{tabular}{|c|c|}
\hline
Parameteraisation & $Y_\nu$  \\
\hline
Complex case ($R\,=\,e^{i {\bf A}}$) &  $\left(
\begin{array}{ccc}
 0.00041\, +0.0008 i & -0.00057+0.00076 i & -0.005-0.0004 i \\
 0.0021\, -0.00091 i & 0.0018\, +0.0013 i & -0.0013+0.013 i \\
 0.0008\, -0.0006 i & 0.0010\, +0.00017 i & 0.0014\, +0.007 i \\
\end{array}
\right)$  \\
\hline
Real case ($R\,=\,e^{ {\bf A}}$)  & $\left(
\begin{array}{ccc}
1.46 \times10^{-6} + 0.00026 i & 1.46 \times 10^{-6} + 0.00023 i &  0.00001 + 0.001 i \\
   7.5 \times10^{-7} + 0.0002 i  & 6.37 \times10^{-7} + 0.00028 i  & 2.57\times10^{-6} + 0.0017 i\\
   6.96\times10^{-7} + 0.00016 i & 5.9 \times10^{-7} + 0.0001 i & 2.39\times10^{-6} + 0.0012 i  \\
  \end{array}
\right)$ \\
\hline
\end{tabular}
\caption{An estimate of the Yukawa coupling order for each case of $R$ matrix forms.  The rows in these matrices indicate corresponding Yukawa coupling associated with three lepton flavors. }
\label{yukawa_table}
\end{center}
\end{table} 

\section{Role of the SSME in the calculation of BR$(\mu \rightarrow e + \gamma )$}\label{lfv}
In the previous sections we mainly discussed about the leptogenesis parameter space in the context of two different choices corresponding to the $R$ matrix. In this section we further investigate, whether the favourable parameter space for leptogenesis can give rise to a desired branching ratio of a LFV decay process. For simplicity we have considered here the branching for $ \mu \rightarrow e + \gamma $ decay process for investigation, which presently provides the strongest bound. As we have seen, the rates of the LFV processes in the canonical type-I seesaw model with massive neutrinos are so strongly suppressed that these processes are not observable in practice, one has {\it e.g.,} BR$(\mu \rightarrow e + \gamma ) < 10^{-47}$ \cite{Cheng:1980tp,Aubert:2009ag}. 

However, the presently planned near future sensitivity of BR$(\mu \rightarrow e + \gamma ) < 6 \times 10^{-14}$, having the present bound  to be $ < 4.2 \times 10^{-13}$ (taken from Refs. \cite{Aubert:2009ag,Adam:2013mnn,Baldini:2013ke}). A search for a theoretically well motivated framework to account for this larger order of magnitude for BR$(\mu \rightarrow e + \gamma )$ is hence looked for. Therefore, we bring forward the required range of the SSME present in the aforementioned $R$ matrix for the type-I seesaw Yukawa coupling which is supposed to bring an enhancement in the BR$(\mu \rightarrow e + \gamma )$. 
\begin{table}[h!]
\begin{center}
\begin{tabular}{|c|c|c|}
\hline
Parameteraisation & ${\rm Br}(\mu\rightarrow e \gamma)$ & {\bf a}  \\
\hline
Complex case ($R\,=\,e^{i {\bf A}}$) &  $5.67 \times 10^{-13}$ & 9.78\\
            & $3.83 \times 10^{-35}$ & 2.42 \\
\hline
Real case ($R\,=\,e^{ {\bf A}}$)  & $2.47 \times 10^{-40}$  & 2.45 \\
\hline
\end{tabular}
\caption{Attainable ${\rm Br}(\mu\rightarrow e \gamma)$ for the complex case and the real case.}
\label{Br_tabdata}
\end{center}
\end{table} 

 The role of the SSME in the connection between lepton asymmetry and branchings of various LFV decays can be understood as follows. The branching ratios of LFV processes and lepton asymmetry are related by the Dirac Yukawa coupling $Y_\nu$. Thus it is advised to look for a common range of the SSME which can simultaneously address a successful leptogenesis and an enhanced branching ratio for the LFV decay channel of our interest.  As we learn, the matrix $Y_\nu$ can be expressed in terms of the light neutrino and heavy RHN masses, the neutrino mixing matrix $U$, and the rotational matrix $R$. Leptogenesis can take place only if $Y_\nu$ is complex. However, this is not a necessary condition for the success of LFV. The branching for a particular LFV process depends explicitly on $(Y_\nu^\dagger Y_\nu)_{ij}$ . Therefore, the predictions on BR($l_i \rightarrow l_j \gamma$) are directly linked to the magnitude of the matrix elements involved in the rotational matrix $R$. Using the complex form of the $R$ matrix authors in \cite{Pascoli:2003rq} have reported two findings. One is that it is difficult to get a common viable regime for leptogenesis and branchings for various LFV decays for such high RHN mass scale and the second is using the complex $R$ matrix as discussed in this analysis the branching ratios for different LFV decays ($l_i \rightarrow l_j \gamma$) are expected to have an unsuppressed value in comparison to what one obtains considering a real $R$ matrix.

\begin{figure*}[h]
\begin{center}
\includegraphics[scale=0.27]{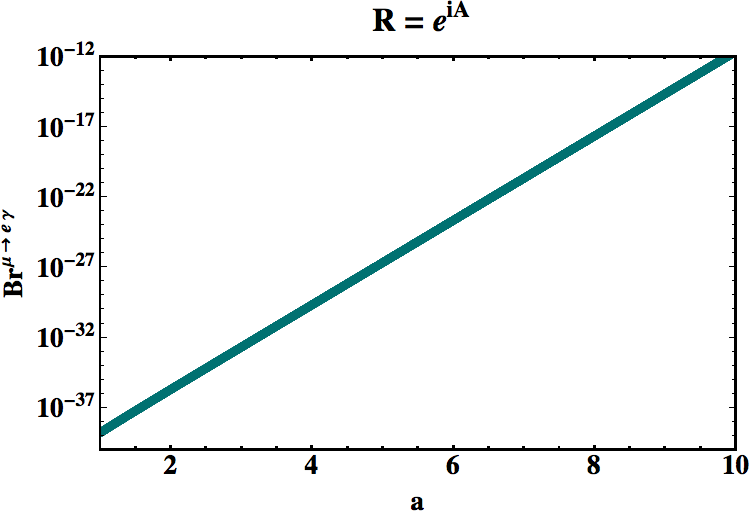}
\includegraphics[scale=0.32]{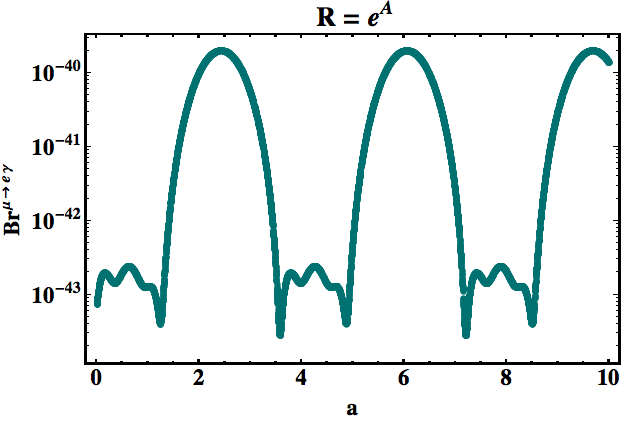}    
\caption{BR($l_i \rightarrow l_j \gamma$) plots as a function of the skew symmetric matrix element for the complex (left) and real (right) cases. }
\label{fig:br}
\end{center}
\end{figure*}

To compute the branching ratio for this decay process we followed prescriptions from \cite{Pascoli:2003rq,Bambhaniya:2016rbb}.
 As illustrated in the previous sections, about the role of the orthogonal matrix being very deterministic in the leptogenesis scenario, that also holds good for the calculation of BR$(\mu \rightarrow e + \gamma)$. This can be understood  from the Table~\ref{Br_tabdata}, where we provide with the requirement of the SSME in order to obtain a desired branching in type-I seesaw scenario. This can also be evident from the Fig.\,\ref{fig:br}, which demonstrates the variation of the branching ratio with respect to {\bf a}.   

In addition to this, we also report that for a large {\bf a} value one could achieve a common regime for both baryon asymmetry and an unsuppressed branching for LFV decays, for a much smaller mass regime for the RHNs in case of the complex $R$ matrix. 
 
 Given the existing experimental bound on ${\rm BR}(\mu \rightarrow e\gamma)$, our results can be used, in particular, to further constrain the SSME parameter space (here, in particular range of {\bf a}) along with a possible mass window for the RHNs starting from a TeV to the one which is canonically required for type-I seesaw scale. This requires a more detailed numerical analysis which is beyond the scope of the present work.

 \subsection{Discussing the results}\label{conclusion1}
The main objective of this work is to investigate the role of the aforementioned forms of the rotational matrix in the context of leptogenesis and lepton flavor violation in typeI seesaw. As a sub objective we have also examined the importance of low energy CP phase, especially the Dirac one in baryogenesis realized through leptogenesis. The first objective is executed through determining the possible range of the SSME for leptogenesis and LFV which leads to the following discussion. 
\begin{itemize}
\item The structural influence provided by the complex $R$ matrix on the neutrino Yukawa couplings leads to important predictions in the low energy parameter space. The complex $R$ matrix imposes a restriction on the Dirac CP phase values, and range of the lightest active neutrino mass in order to account for the observed baryon to photon ratio. These predictions can be further tested in the future and ongoing neutrino oscillation experiments. Not only that, we are able to constrain the SSME ({\bf a}) parameter space in this complex $R$ case, which can further be probed through the Branching ratios of various LFV decays. For the complex $R$ matrix, the connection between low energy and high energy CP violation is found to be very prominent, with a preference for the Dirac CP phase ($\delta$) to be around $\pi/2$. The observed value of the baryon to photon ratio allows the SSME to range from {\bf (1.67 - 2.42)} for the given choice of the RHN mass.

\item For the real $R$ matrix we have noticed an unusual behaviour of the Dirac phase in obtaining a non-zero lepton asymmetry. In order to satisfy the leptogenesis constraint we see that, the real $R$ matrix predicts the Dirac CP phase to be around certain CP conserving values which makes us convenient to write $\delta \approx 0, \, \pi,\, 2\pi$. One can expect for this parameterisation that, since the orthogonal matrix is chosen to be a real one, the source of CP violation in the Yukawa couplings is solely contributed by the low energy CP phase $\delta$, present in the PMNS matrix. Thereby, a CP violating value of $\delta$ can be naturally expected to generate a sufficient amount of lepton asymmetry. However, this is not obtained in the real $R$ case unlike the complex case. For this choice of $R$ matrix we obtain the constrained region of $\delta$ very close to the CP conserving values as mentioned above. This result makes the real case very different from the complex case. Unlike the complex $R$ case, the observed $\eta_B$ restricts the SSME to pick certain values which come after certain periodic intervals. This is due to the reason that the Yukawa couplings in this case are guided by sinusoidal functions of $\sqrt{3}\,{\bf a}$. Some benchmark points however could be noticed like {\bf a = 0.01, 0.35, 1.12} and {\bf 3.65} for the chosen RHN mass appropriate for flavored leptogenesis. However, in both of the two cases of $R$ matrix the required order of the RHN mass degeneracy ($\Delta M$) does not differ much in order to have a successful leptogenesis.

\item  We have shown the influence of the low energy phase on asymmetries associated with each lepton flavor in the form of contour plots. For the complex $R$ case we obtain a direct connection between the maximal low energy CP violation (implying $\delta \approx \pi/2$) and an enhanced lepton asymmetry for all the flavors for any choice of the lightest neutrino mass. This scenario is seemingly different in case of the real $R$ matrix. Such connection between the low and high energy CP violation is realized for the real $R$ matrix for a higher value of the lightest neutrino mass as we notice it to be $0.008$~eV. This correlation does not hold true for a smaller value of the lightest neutrino mass as we see for the case of $m_1 = 0.002$~eV.

\item We find that for successful leptogenesis the required range of the SSME parameter does not bring the required enhancement in the branching ratios of the LFV decay discussed here. 
 Taking into account the leptogenesis constraints on the relevant parameters we learn that an increase of the {\bf a} value can drastically enhance the branching of $\mu \rightarrow e\gamma$ decay approximately by a factor of $10^{12}$ when one considers the complex $R$ matrix for Yukawa extraction. An increase in the branching requires {\bf a} value to be around {\bf 9.78} for an RHN mass of the order $10^8$~GeV. However, this combination is unable to produce an ample amount of baryon asymmetry we are looking for. On the other hand this kind of enhancement in the branching is not at all obtained in the case of the later choice of $R$ matrix, even after choosing a larger value of {\bf a}. This is due to the hyperbolic kind of escalation in the overall Yukawa coupling matrix elements in the complex $R$ case, which is absent in case of the real $R$ matrix. 
\end{itemize}


\section{Conclusion}\label{conclusion2}
In this work we widely explore the notable feature of Casas-Ibarra parameterisation where the rotational matrix has a special form (one being $e^{i {\bf A}}$ and another being $e^{\bf A}$). In the light of these two choices of $R$-matrix we investigate the viable parameter space for flavored leptogenesis in the type-I seesaw. While doing so, the active involvement of these $R$ matrices in the extraction of Dirac Yukawa coupling is realized with the help of CI formalism. The same coupling acts indispensably in the generation of light Majorana neutrino mass, lepton asymmetry and various lepton flavor violating decays. The light neutrino masses are assumed to be offered by the canonical type-I seesaw, with a choice of the RHN mass scale to be around $10^{8}$ GeV along with a quasi degenerate RHN mass spectrum. The results of this particular analysis highlight the reliance of high energy CP violation on the leptonic CP phases present in the form of Dirac CP phase in the PMNS matrix. We noticed that with the given choice of the RHN mass regime, a maximum Dirac CP violation ($\delta\,=\, \pi/2$) can yield the maximum baryon to photon ratio in comparison to what one obtains for other values of $\delta$. Considering the scenario of a wider range of RHN mass has also been explored and found to be less appealing in the context of low and high energy CP connection. However, while working with a varying range of RHN mass scale,  we get an upper bound for the lightest neutrino mass for normal mass ordering,
purely from the baryon asymmetry criteria. It is found to be around $0.015$ eV for both the
choices of the $R$ matrix. This finding can be verified by the ongoing {\bf KATRIN}\,\cite{KATRIN:2021uub} search. It is worth noting that, such bound on the lightest neutrino mass doest not exist for the case where we have fixed the RHN mass at $10^8$~GeV.
The upper bound on the lightest neutrino mass from baryon asymmetry criteria is a notable
feature of this analysis and constitute a new finding for the best of our knowledge. In addition, we have also studied the time evolution of the RHN abundance and flavored asymmetries by numerically solving the coupled Boltzmann equations.

 We have mainly investigated the role of these structures of $R$-matrix in flavored leptogenesis realized only through the Dirac CP violation and attempted to probe the range of SSME through the study of branching ratios of LFV process mainly the 
$\mu \rightarrow e \gamma$ channel with the motivation that, it can constitute an indirect probe of the elements of $R$. There are two main findings of this objective one of them gives us the realization that the leptogenesis parameter space corresponding to the SSME range is far from reaching the present sensitivity of the branching ratio. More specifically the SSME value should be large enough to be testified by the LFV phenomenology. The second one is regarding the correlation between the low and high energy CP violation through leptogenesis. To be precise, for the complex $R$ only such correlation among the low and high energy CP phase can be established for a very narrow regime of the SSME and the $M_R$. Whereas, for the real $R$ we do not get any such notion. The aforementioned finding regarding the prediction on Dirac CP phase for any choice of $R$ do not match when one considers a wider range of the RHN mass along with the SSME and the lightest neutrino mass. Thus the later scenario reflects less predictability in the context of such connection which one can in principle expect.

To conclude, these two special $R$ matrices lead to potentially important phenomenological predictions for some specific choice of parameter space in terms of the SSME and the RHN mass. A detailed investigation is needed to underscore the importance of all the PMNS phases in the leptogenesis process through the consideration of the Casas-Ibarra formalism involving these special choices of the orthogonal matrix, which can be found in an upcoming work. 
A study in a non-minimalistic scenario considering the presence of the Majorana phases, not only can shade light on the source of high energy CP violation but also can exclude some region of free parameter space associated with the effective neutrino mass $m_{\beta \beta}$ governing the neutrino less double beta decay ($0 \nu \beta \beta$) subject to some constraints that might put restrictions on these phases. 

\section*{Acknowledgement}
AM wants to thank Sarbeswar Pal and Abhijit Kumar Saha for having useful discussion. AM would like to acknowledge the financial support provided by SERB-DST, Govt. of India through the project EMR/2017/001434. AM also acknowledges the post doctoral fellowship provided by the Saha Institute of Nuclear Physics, Kolkata. 
\begin{appendices}
\section*{Appendix 1}
\section{Expansion of the $R$ matrices:} \label{appendix_complex}

Following the definition of the skew symmetric matrix ${\bf A}$ from Eq.\,\ref{skew} and the condition of equity among the matrix elements, the characteristic polynomial for ${\bf A}$ can be written as, 
\begin{equation}
|{\bf A} - \lambda I| = 0 \Rightarrow \lambda^3 + \lambda r^2 = 0 ,
\end{equation}
which further leads to, $\lambda^3 = - r^2 \lambda$. Using the Cayley-Hamilton theorem which states that, every square matrix over a commutative ring satisfies its own characteristic equation, one can simply write ${\bf A}^3 = -r^2 {\bf A}$.
Now on expanding, 
$$e^{i{\bf A}} = 1+ i {\bf A} - \frac{{\bf A}^2}{2!}-i  \frac{{\bf A}^3}{3!}+ \frac{\bf{A}^4}{4!}-i  \frac{{\bf A}^5}{5!}+... \rm{etc.}$$

Using ${\bf A}^3 = -r^2 {\bf A}$, and after separating the real and imaginary parts of $e^{i{\bf A}}$ we can write,
\begin{gather*}
\rm Im(e^{i{\bf A}}) = i {\bf A} \frac{1}{r} \left( r+ \frac{r^3}{3!}+.....\right) = i{\bf A} \frac{\sinh r }{r}\\ \nonumber
\rm Re(e^{i{\bf A}}) = - \frac{{\bf A}^2}{r^2} \left(\frac{r^2}{2!}+ \frac{r^4}{4!}+.....\right) = - \frac{{\bf A}^2}{r^2} \left(\cosh r-1\right)
\end{gather*}
The above formulations lead one to write, 
\begin{equation*}
R = e^{i {\bf A}}=1-\frac{\cosh r-1}{r^{2}} {\bf A}^{2}+i\frac{\sinh r}{r} {\bf A}.
\end{equation*}
The same procedure will help us to derive the analogous formula for the other parameterisation when, $R= e^{\bf A}$.

\section*{Appendix 2}
\section{Results for general consideration on $M_R$ scale}\label{appendix2}
\begin{figure*}[t]
\begin{center}
\includegraphics[scale=0.5]{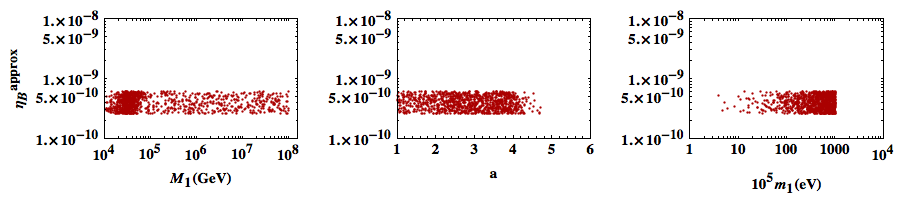}\\
\includegraphics[scale=0.5]{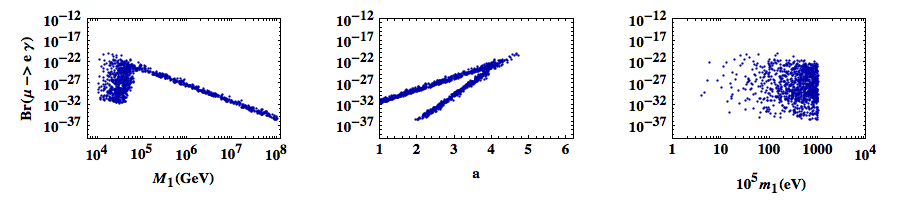}    
\caption{Shows the prediction on the input parameter space with $R= e^{i A}$ having successful Dirac phase induced leptogenesis (top panel) and validating them in the $\mu \rightarrow e \gamma$ searches (bottom panel). In the top presented is the baryon asymmetry parameter as a function of the RHN mass (first), the SSME (second), and the lightest neutrino mass (third). The lower panel provides the information on the branching ratio when plotted against the aforementioned parameters.}
\label{fig:ghost1}
\end{center}
\end{figure*}

\begin{figure*}[h]
\begin{center}
\includegraphics[scale=0.23]{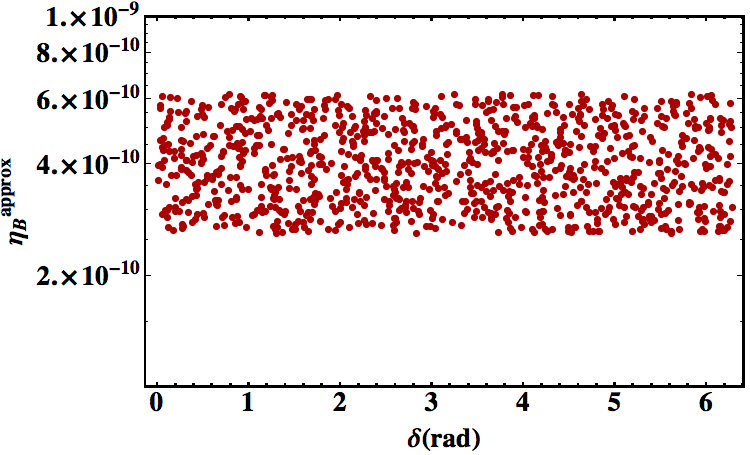}
\includegraphics[scale=0.23]{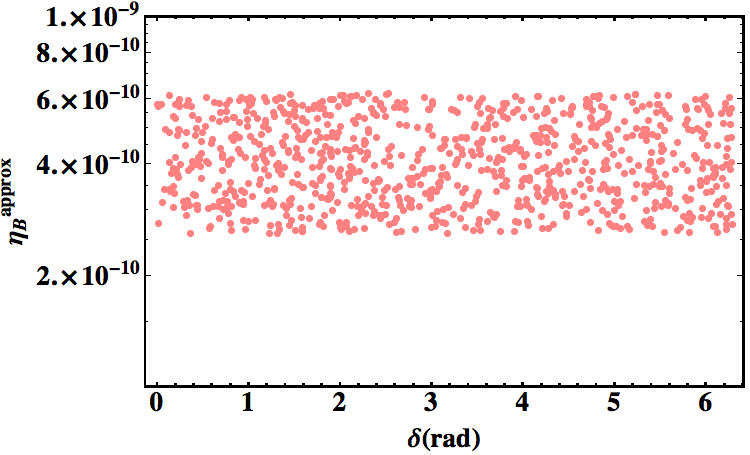}    
\caption{Shows the reliance of the baryon asymmetry parameter on the Dirac CP phase for $R =  e^{i{\bf A}}$ (left) and $R =  e^{\bf A}$ (right).}
\label{fig:ghost3}
\end{center}
\end{figure*}

In this section we provide the viability test for leptogenesis with the present theoretical background for a random scan over the parameters $M_R$, ${\bf a}$, $\delta$(as before) and $m_1$. Then we numerically evaluate the baryon asymmetry with the new set of parameter choices comprizing of $a = 1\,-\,6$, the lightest neutrino mass $m_1 = 10^{-5}\,-\,0.01$~eV along with the RHN mass $M_R=10^4$ to $10^8$ GeV. We found that the $\eta_B$ criteria along with $\epsilon \leq 1$ together impose constraints on the input parameter space, mainly which is associated with the SSME values. We use these left over parameter space for the calculation of the Br$(\mu \rightarrow e + \gamma)$ to check the maximum branching ratio.  For the complex case of $R$ ($R\,=\,e^{i\bf A}$), the results are demonstrated in the Fig.~\ref{fig:ghost1}. From the top panel of this figure it is noticed that, there exists an upper bound for the SSME which is 4.8 and found to be uniform for any mass scale within the chosen $M_N$ range. It is to mention that, larger values of {\bf a} may lead to very high Yukawa coupling making the lepton asymmetry ($\epsilon_1 $) of order more than 1. In the second figure of the top panel, the blank region after $a= 4.8$ yields  $\epsilon_1 > 1$, which led to exclusion of the SSME parameter space. The $\eta_B$ criteria sets a lower bound also on the lightest neutrino mass which is around $10^{-4}$eV, taking a preference towards the larger values of the $m_1$ range (around $10^{-2}$\,eV). This upper bound on the $m_{\rm lightest}$ can be verified in the light of ongoing {\bf KATRIN}\,\cite{KATRIN:2021uub} searches which recently has reported the upper bound on the effective electron anti-neutrino mass to be $< 0.8$\,eV. Also, this finding on the $m_{\rm lightest}$ from the requirement of successful leptogenesis is new for the best of our knowledge. In the bottom panel we present the branching ratio obtained for the  $\mu \rightarrow e \gamma$ process using the parameter space viable for leptogenesis. The first figure tells us that the branching ratio is a decreasing function of $M_R$ scale. This is trivially understandable from the fact that with increasing $M_R$ the light heavy mixing governing this LFV decay becomes smaller, since this mixing is determined by the ratio $m_D / M_R$. However, it is evident from the first two figures of the bottom panel that the maximum branching ($\mathcal{O} (10^{-21})$) is reached  for a quite smaller RHN mass (around $\mathcal{O} (10^{4})$)  and essentially accompanied by $a\,=\,4.8$. The lightest neutrino mass does not play such decisive role in the order of magnitude rise/fall of the branching ratio as evident in the third figure of the bottom panel. Another important finding is regarding the Dirac CP dependency of the baryon asymmetry in such a scenario of general scan over these parameters. We do not have any preference over a particular range/value of $\delta$ when one considers a wide range of the RHN mass and the SSME, as depicted in Fig.~\ref{fig:ghost3}. From this explanation one can be clear about two findings, one is the leptogenesis parameter space with the first choice of $R$ is difficult to be tested in the respective $\mu \rightarrow e \gamma$ searches. The second finding is regarding the less predictive correlation between the low energy CP phase and leptogenesis, which we obtained for a fixed $M_R$ and $a$ (see Fig. \ref{eiA_low}).

\begin{figure*}[h]
\begin{center}
\includegraphics[scale=0.5]{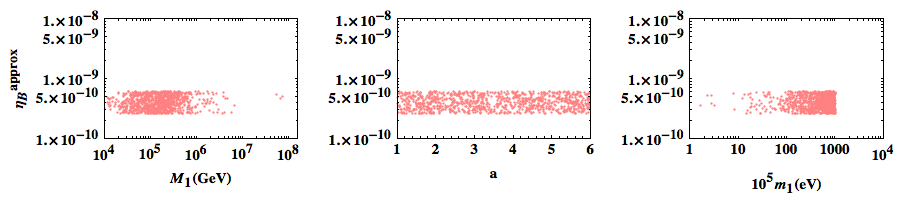}\\
\includegraphics[scale=0.5]{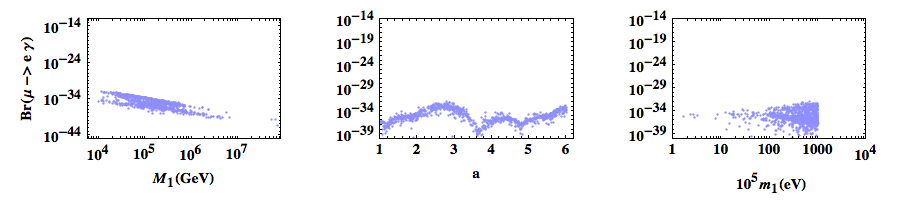}    
\caption{Same as of the caption of Fig.\ref{fig:ghost2}, except for the case with $R =  e^{A}$}
\label{fig:ghost2}
\end{center}
\end{figure*}
\begin{figure*}
\begin{center}
\includegraphics[scale=0.3]{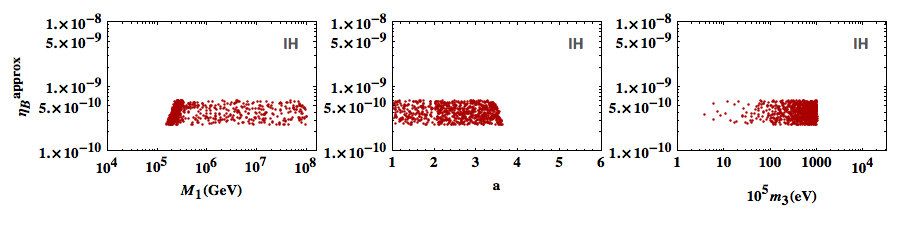}\\ 
\includegraphics[scale=0.5]{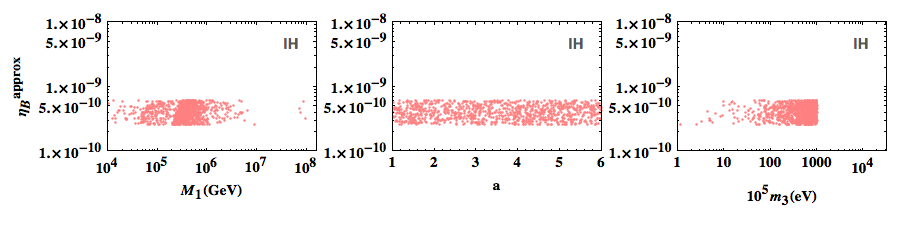}
\caption{Shows the overall result for $\eta_B$ w.r.t. the RHN mass, the SSME and the lightest neutrino mass considering the IH mass pattern of neutrinos for the complex $R$ case (top panel) and real $R$ case (bottom panel). }
\label{fig:IH_complex_real}
\end{center}
\end{figure*}

In Fig.\ref{fig:ghost2} we present the phenomenology corresponding to the case when $R = e^{\bf A}$. In the top panel of this figure it is evident that, the RHN mass scale gets restriction from the $\eta_B$ criteria having an upper bound at around $10^7$ GeV\footnote{although a few points are visible at around $M_R = 10^8$GeV.}. Whereas the SSME remains completely unaffected by the $\eta_B$ constraint. The reason being the periodic nature (rather being hyperbolic increasing) of the Yukawa coupling w.r.t. the parameter {\bf a}. The prediction on the lightest neutrino mass also remains the same with the previous case of the $R$ matrix. On the other hand the maximum branching ratio obtained with the later choice of $R$ with the left over parameter space is $\mathcal{O}(10^{-31})$, which is quite far from any present or future sensitivity of $\mu \rightarrow e \gamma$ searches. From the branching ratio requirement the former choice serves better as it yields quite a large order of magnitude. However, the expectation of correlation between low and high energy CP violation is not encouraged for such a general scan over the input parameters. We have performed similar exercise for the inverted hierarchy of light neutrino mass the results of which have been shown in Fig.~\ref{fig:IH_complex_real}. It is evident from this figure that, the change of mass hierarchy does not alter the baryon asymmetry parameter space related to the SSME parameter space and the lightest neutrino mass. However, a lower bound for the RHN mass is imposed by the baryon asymmetry criteria for the IH of light neutrino mass. We do not present the branching ratio results for the IH case, as there is no significant change in the branching ratio observed.

\end{appendices}
\pagebreak
\pagebreak
\bibliography{leptogenesis}

\providecommand{\href}[2]{#2}\begingroup\raggedright\begin{thebibliography}{10}

\bibitem{Fukuda:2001nk}
{\bf Super-Kamiokande} Collaboration, S.~Fukuda et~al., {\it {Constraints on
  neutrino oscillations using 1258 days of Super-Kamiokande solar neutrino
  data}},  {\em Phys. Rev. Lett.} {\bf 86} (2001) 5656--5660,
  [\href{http://arxiv.org/abs/hep-ex/0103033}{{\tt hep-ex/0103033}}].

\bibitem{Ahmad:2002jz}
{\bf SNO} Collaboration, Q.~R. Ahmad et~al., {\it {Direct evidence for neutrino
  flavor transformation from neutral current interactions in the Sudbury
  Neutrino Observatory}},  {\em Phys. Rev. Lett.} {\bf 89} (2002) 011301,
  [\href{http://arxiv.org/abs/nucl-ex/0204008}{{\tt nucl-ex/0204008}}].

\bibitem{Ahmad:2002ka}
{\bf SNO} Collaboration, Q.~R. Ahmad et~al., {\it {Measurement of day and night
  neutrino energy spectra at SNO and constraints on neutrino mixing
  parameters}},  {\em Phys. Rev. Lett.} {\bf 89} (2002) 011302,
  [\href{http://arxiv.org/abs/nucl-ex/0204009}{{\tt nucl-ex/0204009}}].

\bibitem{Eguchi:2002dm}
{\bf KamLAND} Collaboration, K.~Eguchi et~al., {\it {First results from
  KamLAND: Evidence for reactor anti-neutrino disappearance}},  {\em Phys. Rev.
  Lett.} {\bf 90} (2003) 021802,
  [\href{http://arxiv.org/abs/hep-ex/0212021}{{\tt hep-ex/0212021}}].

\bibitem{Abe:2008aa}
{\bf KamLAND} Collaboration, S.~Abe et~al., {\it {Precision Measurement of
  Neutrino Oscillation Parameters with KamLAND}},  {\em Phys. Rev. Lett.} {\bf
  100} (2008) 221803, [\href{http://arxiv.org/abs/0801.4589}{{\tt
  arXiv:0801.4589}}].

\bibitem{Abe:2011sj}
{\bf T2K} Collaboration, K.~Abe et~al., {\it {Indication of Electron Neutrino
  Appearance from an Accelerator-produced Off-axis Muon Neutrino Beam}},  {\em
  Phys. Rev. Lett.} {\bf 107} (2011) 041801,
  [\href{http://arxiv.org/abs/1106.2822}{{\tt arXiv:1106.2822}}].

\bibitem{Abe:2011fz}
{\bf Double Chooz} Collaboration, Y.~Abe et~al., {\it {Indication of Reactor
  $\bar{\nu}_e$ Disappearance in the Double Chooz Experiment}},  {\em Phys.
  Rev. Lett.} {\bf 108} (2012) 131801,
  [\href{http://arxiv.org/abs/1112.6353}{{\tt arXiv:1112.6353}}].

\bibitem{An:2012eh}
{\bf Daya Bay} Collaboration, F.~P. An et~al., {\it {Observation of
  electron-antineutrino disappearance at Daya Bay}},  {\em Phys. Rev. Lett.}
  {\bf 108} (2012) 171803, [\href{http://arxiv.org/abs/1203.1669}{{\tt
  arXiv:1203.1669}}].

\bibitem{Ahn:2012nd}
{\bf RENO} Collaboration, J.~K. Ahn et~al., {\it {Observation of Reactor
  Electron Antineutrino Disappearance in the RENO Experiment}},  {\em Phys.
  Rev. Lett.} {\bf 108} (2012) 191802,
  [\href{http://arxiv.org/abs/1204.0626}{{\tt arXiv:1204.0626}}].

\bibitem{Fukugita:1986hr}
M.~Fukugita and T.~Yanagida, {\it {Baryogenesis Without Grand Unification}},
  {\em Phys. Lett.} {\bf B174} (1986) 45--47.

\bibitem{Hinshaw:2012aka}
{\bf WMAP} Collaboration, G.~Hinshaw et~al., {\it {Nine-Year Wilkinson
  Microwave Anisotropy Probe (WMAP) Observations: Cosmological Parameter
  Results}},  {\em Astrophys. J. Suppl.} {\bf 208} (2013) 19,
  [\href{http://arxiv.org/abs/1212.5226}{{\tt arXiv:1212.5226}}].

\bibitem{Aghanim:2018eyx}
{\bf Planck} Collaboration, N.~Aghanim et~al., {\it {Planck 2018 results. VI.
  Cosmological parameters}},  {\em Astron. Astrophys.} {\bf 641} (2020) A6,
  [\href{http://arxiv.org/abs/1807.06209}{{\tt arXiv:1807.06209}}].

\bibitem{ade2016planck}
P.~A. Ade, N.~Aghanim, M.~Arnaud, M.~Ashdown, J.~Aumont, C.~Baccigalupi,
  A.~Banday, R.~Barreiro, J.~Bartlett, N.~Bartolo, et~al., {\it Planck 2015
  results-xiii. cosmological parameters},  {\em Astronomy \& Astrophysics} {\bf
  594} (2016) A13.

\bibitem{Minkowski:1977sc}
P.~Minkowski, {\it {$\mu \to e\gamma$ at a Rate of One Out of $10^{9}$ Muon
  Decays?}},  {\em Phys. Lett. B} {\bf 67} (1977) 421--428.

\bibitem{Mohapatra:1979ia}
R.~N. Mohapatra and G.~Senjanovic, {\it {Neutrino Mass and Spontaneous Parity
  Nonconservation}},  {\em Phys. Rev. Lett.} {\bf 44} (1980) 912.

\bibitem{Ellis:1992nq}
J.~R. Ellis, D.~V. Nanopoulos, and K.~A. Olive, {\it {Flipped heavy neutrinos:
  From the solar neutrino problem to baryogenesis}},  {\em Phys. Lett. B} {\bf
  300} (1993) 121--127, [\href{http://arxiv.org/abs/hep-ph/9211325}{{\tt
  hep-ph/9211325}}].

\bibitem{Rubakov:1996vz}
V.~Rubakov and M.~Shaposhnikov, {\it {Electroweak baryon number nonconservation
  in the early universe and in high-energy collisions}},  {\em Usp. Fiz. Nauk}
  {\bf 166} (1996) 493--537, [\href{http://arxiv.org/abs/hep-ph/9603208}{{\tt
  hep-ph/9603208}}].

\bibitem{PhysRevD.49.6394}
J.~M. Cline, K.~Kainulainen, and K.~A. Olive, {\it Protecting the primordial
  baryon asymmetry from erasure by sphalerons},  {\em Phys. Rev. D} {\bf 49}
  (Jun, 1994) 6394--6409.

\bibitem{DOnofrio:2012phz}
M.~D'Onofrio, K.~Rummukainen, and A.~Tranberg, {\it {The Sphaleron Rate through
  the Electroweak Cross-over}},  {\em JHEP} {\bf 08} (2012) 123,
  [\href{http://arxiv.org/abs/1207.0685}{{\tt arXiv:1207.0685}}].

\bibitem{Magg:1980ut}
M.~Magg and C.~Wetterich, {\it {Neutrino Mass Problem and Gauge Hierarchy}},
  {\em Phys. Lett. B} {\bf 94} (1980) 61--64.

\bibitem{Schechter:1980gr}
J.~Schechter and J.~W.~F. Valle, {\it {Neutrino Masses in SU(2) x U(1)
  Theories}},  {\em Phys. Rev. D} {\bf 22} (1980) 2227.

\bibitem{Ma:1998dn}
E.~Ma, {\it {Pathways to naturally small neutrino masses}},  {\em Phys. Rev.
  Lett.} {\bf 81} (1998) 1171--1174,
  [\href{http://arxiv.org/abs/hep-ph/9805219}{{\tt hep-ph/9805219}}].

\bibitem{Barbieri:1999ma}
R.~Barbieri, P.~Creminelli, A.~Strumia, and N.~Tetradis, {\it {Baryogenesis
  through leptogenesis}},  {\em Nucl. Phys. B} {\bf 575} (2000) 61--77,
  [\href{http://arxiv.org/abs/hep-ph/9911315}{{\tt hep-ph/9911315}}].

\bibitem{Buchmuller:1999cu}
W.~Buchmuller and M.~Plumacher, {\it {Matter antimatter asymmetry and neutrino
  properties}},  {\em Phys. Rept.} {\bf 320} (1999) 329--339,
  [\href{http://arxiv.org/abs/hep-ph/9904310}{{\tt hep-ph/9904310}}].

\bibitem{Buchmuller:2003jr}
W.~Buchmuller, {\it {Neutrinos and matter antimatter asymmetry of the
  universe}},  in {\em {10th International Workshop on Neutrino Telescopes}},
  6, 2003.
\newblock \href{http://arxiv.org/abs/hep-ph/0306047}{{\tt hep-ph/0306047}}.

\bibitem{Nardi:2006fx}
E.~Nardi, Y.~Nir, E.~Roulet, and J.~Racker, {\it {The Importance of flavor in
  leptogenesis}},  {\em JHEP} {\bf 01} (2006) 164,
  [\href{http://arxiv.org/abs/hep-ph/0601084}{{\tt hep-ph/0601084}}].

\bibitem{Rahat:2020mio}
M.~H. Rahat, {\it {Leptogenesis from the Asymmetric Texture}},  {\em Phys. Rev.
  D} {\bf 103} (2021) 035011, [\href{http://arxiv.org/abs/2008.04204}{{\tt
  arXiv:2008.04204}}].

\bibitem{Dev:2017trv}
P.~S.~B. Dev, P.~Di~Bari, B.~Garbrecht, S.~Lavignac, P.~Millington, and
  D.~Teresi, {\it {Flavor effects in leptogenesis}},  {\em Int. J. Mod. Phys.
  A} {\bf 33} (2018) 1842001, [\href{http://arxiv.org/abs/1711.02861}{{\tt
  arXiv:1711.02861}}].

\bibitem{Blanchet:2006be}
S.~Blanchet and P.~Di~Bari, {\it {Flavor effects on leptogenesis predictions}},
   {\em JCAP} {\bf 03} (2007) 018,
  [\href{http://arxiv.org/abs/hep-ph/0607330}{{\tt hep-ph/0607330}}].

\bibitem{Mukherjee:2018fms}
A.~Mukherjee, M.~K. Das, and J.~K. Sarma, {\it {Normal hierarchy neutrino mass
  model revisited with leptogenesis}},
  \href{http://arxiv.org/abs/1803.08239}{{\tt arXiv:1803.08239}}.

\bibitem{Narendra:2020hoz}
N.~Narendra, N.~Sahu, and S.~Uma~Sankar, {\it {Flavoured CP-asymmetry at the
  effective neutrino mass floor}},  {\em Nucl. Phys. B} {\bf 962} (2021)
  115268, [\href{http://arxiv.org/abs/2002.08753}{{\tt arXiv:2002.08753}}].

\bibitem{Buchmuller:2004nz}
W.~Buchmuller, P.~Di~Bari, and M.~Plumacher, {\it {Leptogenesis for
  pedestrians}},  {\em Annals Phys.} {\bf 315} (2005) 305--351,
  [\href{http://arxiv.org/abs/hep-ph/0401240}{{\tt hep-ph/0401240}}].

\bibitem{Giudice:2003jh}
G.~Giudice, A.~Notari, M.~Raidal, A.~Riotto, and A.~Strumia, {\it {Towards a
  complete theory of thermal leptogenesis in the SM and MSSM}},  {\em Nucl.
  Phys. B} {\bf 685} (2004) 89--149,
  [\href{http://arxiv.org/abs/hep-ph/0310123}{{\tt hep-ph/0310123}}].

\bibitem{Davidson:2008bu}
S.~Davidson, E.~Nardi, and Y.~Nir, {\it {Leptogenesis}},  {\em Phys. Rept.}
  {\bf 466} (2008) 105--177, [\href{http://arxiv.org/abs/0802.2962}{{\tt
  arXiv:0802.2962}}].

\bibitem{Plumacher:1996kc}
M.~Plumacher, {\it {Baryogenesis and lepton number violation}},  {\em Z. Phys.
  C} {\bf 74} (1997) 549--559, [\href{http://arxiv.org/abs/hep-ph/9604229}{{\tt
  hep-ph/9604229}}].

\bibitem{Buchmuller:2004tu}
W.~Buchmuller, P.~Di~Bari, and M.~Plumacher, {\it {Some aspects of thermal
  leptogenesis}},  {\em New J. Phys.} {\bf 6} (2004) 105,
  [\href{http://arxiv.org/abs/hep-ph/0406014}{{\tt hep-ph/0406014}}].

\bibitem{Pilaftsis:2003gt}
A.~Pilaftsis and T.~E. Underwood, {\it {Resonant leptogenesis}},  {\em Nucl.
  Phys. B} {\bf 692} (2004) 303--345,
  [\href{http://arxiv.org/abs/hep-ph/0309342}{{\tt hep-ph/0309342}}].

\bibitem{Deppisch:2010fr}
F.~F. Deppisch and A.~Pilaftsis, {\it {Lepton Flavour Violation and theta(13)
  in Minimal Resonant Leptogenesis}},  {\em Phys. Rev. D} {\bf 83} (2011)
  076007, [\href{http://arxiv.org/abs/1012.1834}{{\tt arXiv:1012.1834}}].

\bibitem{Casas:2001sr}
J.~A. Casas and A.~Ibarra, {\it {Oscillating neutrinos and $\mu \to e,
  \gamma$}},  {\em Nucl. Phys. B} {\bf 618} (2001) 171--204,
  [\href{http://arxiv.org/abs/hep-ph/0103065}{{\tt hep-ph/0103065}}].

\bibitem{Pascoli:2003rq}
S.~Pascoli, S.~T. Petcov, and C.~E. Yaguna, {\it {Quasidegenerate neutrino mass
  spectrum, mu ---\ensuremath{>} e + gamma decay and leptogenesis}},  {\em
  Phys. Lett. B} {\bf 564} (2003) 241--254,
  [\href{http://arxiv.org/abs/hep-ph/0301095}{{\tt hep-ph/0301095}}].

\bibitem{Berger:1999bg}
M.~S. Berger and B.~Brahmachari, {\it {Leptogenesis and Yukawa textures}},
  {\em Phys. Rev. D} {\bf 60} (1999) 073009,
  [\href{http://arxiv.org/abs/hep-ph/9903406}{{\tt hep-ph/9903406}}].

\bibitem{Joshipura:1999is}
A.~S. Joshipura and E.~A. Paschos, {\it {Constraining leptogenesis from
  laboratory experiments}},  \href{http://arxiv.org/abs/hep-ph/9906498}{{\tt
  hep-ph/9906498}}.

\bibitem{Falcone:2000ib}
D.~Falcone and F.~Tramontano, {\it {Leptogenesis and neutrino parameters}},
  {\em Phys. Rev. D} {\bf 63} (2001) 073007,
  [\href{http://arxiv.org/abs/hep-ph/0011053}{{\tt hep-ph/0011053}}].

\bibitem{Joshipura:2001ui}
A.~S. Joshipura, E.~A. Paschos, and W.~Rodejohann, {\it {A Simple connection
  between neutrino oscillation and leptogenesis}},  {\em JHEP} {\bf 08} (2001)
  029, [\href{http://arxiv.org/abs/hep-ph/0105175}{{\tt hep-ph/0105175}}].

\bibitem{Endoh:2000hc}
T.~Endoh, T.~Morozumi, T.~Onogi, and A.~Purwanto, {\it {CP violation in seesaw
  model}},  {\em Phys. Rev. D} {\bf 64} (2001) 013006,
  [\href{http://arxiv.org/abs/hep-ph/0012345}{{\tt hep-ph/0012345}}]. [Erratum:
  Phys.Rev.D 64, 059904 (2001)].

\bibitem{Rebelo:2002wj}
M.~N. Rebelo, {\it {Leptogenesis without CP violation at low-energies}},  {\em
  Phys. Rev. D} {\bf 67} (2003) 013008,
  [\href{http://arxiv.org/abs/hep-ph/0207236}{{\tt hep-ph/0207236}}].

\bibitem{Branco:2002kt}
G.~C. Branco, R.~Gonzalez~Felipe, F.~R. Joaquim, and M.~N. Rebelo, {\it
  {Leptogenesis, CP violation and neutrino data: What can we learn?}},  {\em
  Nucl. Phys. B} {\bf 640} (2002) 202--232,
  [\href{http://arxiv.org/abs/hep-ph/0202030}{{\tt hep-ph/0202030}}].

\bibitem{Ellis:2002xg}
J.~R. Ellis and M.~Raidal, {\it {Leptogenesis and the violation of lepton
  number and CP at low-energies}},  {\em Nucl. Phys. B} {\bf 643} (2002)
  229--246, [\href{http://arxiv.org/abs/hep-ph/0206174}{{\tt hep-ph/0206174}}].

\bibitem{Frampton:2002qc}
P.~H. Frampton, S.~L. Glashow, and T.~Yanagida, {\it {Cosmological sign of
  neutrino CP violation}},  {\em Phys. Lett. B} {\bf 548} (2002) 119--121,
  [\href{http://arxiv.org/abs/hep-ph/0208157}{{\tt hep-ph/0208157}}].

\bibitem{Endoh:2002wm}
T.~Endoh, S.~Kaneko, S.~K. Kang, T.~Morozumi, and M.~Tanimoto, {\it {CP
  violation in neutrino oscillation and leptogenesis}},  {\em Phys. Rev. Lett.}
  {\bf 89} (2002) 231601, [\href{http://arxiv.org/abs/hep-ph/0209020}{{\tt
  hep-ph/0209020}}].

\bibitem{Rodejohann:2002hx}
W.~Rodejohann, {\it {Leptogenesis, mass hierarchies and low-energy
  parameters}},  {\em Phys. Lett. B} {\bf 542} (2002) 100--110,
  [\href{http://arxiv.org/abs/hep-ph/0207053}{{\tt hep-ph/0207053}}].

\bibitem{Davidson:2002em}
S.~Davidson and A.~Ibarra, {\it {Leptogenesis and low-energy phases}},  {\em
  Nucl. Phys. B} {\bf 648} (2003) 345--375,
  [\href{http://arxiv.org/abs/hep-ph/0206304}{{\tt hep-ph/0206304}}].

\bibitem{Pascoli:2003uh}
S.~Pascoli, S.~T. Petcov, and W.~Rodejohann, {\it {On the connection of
  leptogenesis with low-energy CP violation and LFV charged lepton decays}},
  {\em Phys. Rev. D} {\bf 68} (2003) 093007,
  [\href{http://arxiv.org/abs/hep-ph/0302054}{{\tt hep-ph/0302054}}].

\bibitem{Molinaro:2008rg}
E.~Molinaro and S.~T. Petcov, {\it {The Interplay Between the 'Low' and 'High'
  Energy CP-Violation in Leptogenesis}},  {\em Eur. Phys. J. C} {\bf 61} (2009)
  93--109, [\href{http://arxiv.org/abs/0803.4120}{{\tt arXiv:0803.4120}}].

\bibitem{Moffat:2018smo}
K.~Moffat, S.~Pascoli, S.~T. Petcov, and J.~Turner, {\it {Leptogenesis from Low
  Energy $CP$ Violation}},  {\em JHEP} {\bf 03} (2019) 034,
  [\href{http://arxiv.org/abs/1809.08251}{{\tt arXiv:1809.08251}}].

\bibitem{Li:2021tlv}
S.-P. Li, X.-Q. Li, X.-S. Yan, and Y.-D. Yang, {\it {Baryogenesis from
  Hierarchical Dirac Neutrinos}},  \href{http://arxiv.org/abs/2105.01317}{{\tt
  arXiv:2105.01317}}.

\bibitem{Pascoli:2006ci}
S.~Pascoli, S.~T. Petcov, and A.~Riotto, {\it {Leptogenesis and Low Energy CP
  Violation in Neutrino Physics}},  {\em Nucl. Phys. B} {\bf 774} (2007) 1--52,
  [\href{http://arxiv.org/abs/hep-ph/0611338}{{\tt hep-ph/0611338}}].

\bibitem{Arias-Aragon:2022ats}
F.~Arias-Arag\'on, E.~Fern\'andez-Mart\'\i{}nez, M.~Gonz\'alez-L\'opez, and
  L.~Merlo, {\it {Dynamical Minimal Flavour Violating inverse seesaw}},  {\em
  JHEP} {\bf 09} (2022) 210, [\href{http://arxiv.org/abs/2204.04672}{{\tt
  arXiv:2204.04672}}].

\bibitem{Abe:2019vii}
{\bf T2K} Collaboration, K.~Abe et~al., {\it {Constraint on the
  matter\textendash{}antimatter symmetry-violating phase in neutrino
  oscillations}},  {\em Nature} {\bf 580} (2020), no.~7803 339--344,
  [\href{http://arxiv.org/abs/1910.03887}{{\tt arXiv:1910.03887}}]. [Erratum:
  Nature 583, E16 (2020)].

\bibitem{Davidson:2007va}
S.~Davidson, J.~Garayoa, F.~Palorini, and N.~Rius, {\it {Insensitivity of
  flavoured leptogenesis to low energy CP violation}},  {\em Phys. Rev. Lett.}
  {\bf 99} (2007) 161801, [\href{http://arxiv.org/abs/0705.1503}{{\tt
  arXiv:0705.1503}}].

\bibitem{Davidson:2004wi}
S.~Davidson, {\it {Parametrizations of the seesaw, or, can the seesaw be
  tested?}},  in {\em {SEESAW25: International Conference on the Seesaw
  Mechanism and the Neutrino Mass}}, pp.~249--260, 9, 2004.
\newblock \href{http://arxiv.org/abs/hep-ph/0409339}{{\tt hep-ph/0409339}}.

\bibitem{Petcov:2006pc}
S.~T. Petcov and T.~Shindou, {\it {Charged lepton decays l(i) ---\ensuremath{>}
  l(j) + gamma, leptogenesis CP-violating parameters and Majorana phases}},
  {\em Phys. Rev. D} {\bf 74} (2006) 073006,
  [\href{http://arxiv.org/abs/hep-ph/0605151}{{\tt hep-ph/0605151}}].

\bibitem{Ibarra:2003up}
A.~Ibarra and G.~G. Ross, {\it {Neutrino phenomenology: The Case of two
  right-handed neutrinos}},  {\em Phys. Lett. B} {\bf 591} (2004) 285--296,
  [\href{http://arxiv.org/abs/hep-ph/0312138}{{\tt hep-ph/0312138}}].

\bibitem{Xing:2009vb}
Z.-z. Xing, {\it {Casas-Ibarra Parametrization and Unflavored Leptogenesis}},
  {\em Chin. Phys. C} {\bf 34} (2010) 1--6,
  [\href{http://arxiv.org/abs/0902.2469}{{\tt arXiv:0902.2469}}].

\bibitem{Chakraborty:2020gqc}
M.~Chakraborty, R.~Krishnan, and A.~Ghosal, {\it {Predictive $S_4$ flavon model
  with $\text{TM}_1$ mixing and baryogenesis through leptogenesis}},  {\em
  JHEP} {\bf 09} (2020) 025, [\href{http://arxiv.org/abs/2003.00506}{{\tt
  arXiv:2003.00506}}].

\bibitem{Xing:2020erm}
Z.-z. Xing and D.~Zhang, {\it {A direct link between unflavored leptogenesis
  and low-energy CP violation via the one-loop quantum corrections}},  {\em
  JHEP} {\bf 04} (2020) 179, [\href{http://arxiv.org/abs/2003.00480}{{\tt
  arXiv:2003.00480}}].

\bibitem{Borah:2020wyc}
D.~Borah, S.~Jyoti~Das, and A.~K. Saha, {\it {Cosmic inflation in minimal
  $U(1)_{B-L}$ model: implications for (non) thermal dark matter and
  leptogenesis}},  {\em Eur. Phys. J. C} {\bf 81} (2021), no.~2 169,
  [\href{http://arxiv.org/abs/2005.11328}{{\tt arXiv:2005.11328}}].

\bibitem{Esteban:2020cvm}
I.~Esteban, M.~C. Gonzalez-Garcia, M.~Maltoni, T.~Schwetz, and A.~Zhou, {\it
  {The fate of hints: updated global analysis of three-flavor neutrino
  oscillations}},  {\em JHEP} {\bf 09} (2020) 178,
  [\href{http://arxiv.org/abs/2007.14792}{{\tt arXiv:2007.14792}}].

\bibitem{Branco:1998bw}
G.~C. Branco, M.~N. Rebelo, and J.~I. Silva-Marcos, {\it {Degenerate and
  quasidegenerate Majorana neutrinos}},  {\em Phys. Rev. Lett.} {\bf 82} (1999)
  683--686, [\href{http://arxiv.org/abs/hep-ph/9810328}{{\tt hep-ph/9810328}}].

\bibitem{Petcov:2005yh}
S.~T. Petcov, T.~Shindou, and Y.~Takanishi, {\it {Majorana CP-violating phases,
  RG running of neutrino mixing parameters and charged lepton flavor violating
  decays}},  {\em Nucl. Phys. B} {\bf 738} (2006) 219--242,
  [\href{http://arxiv.org/abs/hep-ph/0508243}{{\tt hep-ph/0508243}}].

\bibitem{Konar:2020vuu}
P.~Konar, A.~Mukherjee, A.~K. Saha, and S.~Show, {\it {A dark clue to seesaw
  and leptogenesis in a pseudo-Dirac singlet doublet scenario with
  (non)standard cosmology}},  \href{http://arxiv.org/abs/2007.15608}{{\tt
  arXiv:2007.15608}}.

\bibitem{Dev:2015wpa}
P.~S.~B. Dev, P.~Millington, A.~Pilaftsis, and D.~Teresi, {\it {Corrigendum to
  ''Flavour Covariant Transport Equations: an Application to Resonant
  Leptogenesis''}},  {\em Nucl. Phys. B} {\bf 897} (2015) 749--756,
  [\href{http://arxiv.org/abs/1504.07640}{{\tt arXiv:1504.07640}}].

\bibitem{Abada:2006ea}
A.~Abada, S.~Davidson, A.~Ibarra, F.~X. Josse-Michaux, M.~Losada, and
  A.~Riotto, {\it {Flavour Matters in Leptogenesis}},  {\em JHEP} {\bf 09}
  (2006) 010, [\href{http://arxiv.org/abs/hep-ph/0605281}{{\tt
  hep-ph/0605281}}].

\bibitem{Adhikary:2014qba}
B.~Adhikary, M.~Chakraborty, and A.~Ghosal, {\it {Flavored leptogenesis with
  quasidegenerate neutrinos in a broken cyclic symmetric model}},  {\em Phys.
  Rev. D} {\bf 93} (2016), no.~11 113001,
  [\href{http://arxiv.org/abs/1407.6173}{{\tt arXiv:1407.6173}}].

\bibitem{Dolan:2018qpy}
M.~J. Dolan, T.~P. Dutka, and R.~R. Volkas, {\it {Dirac-Phase Thermal
  Leptogenesis in the extended Type-I Seesaw Model}},  {\em JCAP} {\bf 06}
  (2018) 012, [\href{http://arxiv.org/abs/1802.08373}{{\tt arXiv:1802.08373}}].

\bibitem{Bambhaniya:2016rbb}
G.~Bambhaniya, P.~S. Bhupal~Dev, S.~Goswami, S.~Khan, and W.~Rodejohann, {\it
  {Naturalness, Vacuum Stability and Leptogenesis in the Minimal Seesaw
  Model}},  {\em Phys. Rev. D} {\bf 95} (2017), no.~9 095016,
  [\href{http://arxiv.org/abs/1611.03827}{{\tt arXiv:1611.03827}}].

\bibitem{Capozzi:2018ubv}
F.~Capozzi, E.~Lisi, A.~Marrone, and A.~Palazzo, {\it {Current unknowns in the
  three neutrino framework}},  {\em Prog. Part. Nucl. Phys.} {\bf 102} (2018)
  48--72, [\href{http://arxiv.org/abs/1804.09678}{{\tt arXiv:1804.09678}}].

\bibitem{Cheng:1980tp}
T.~P. Cheng and L.-F. Li, {\it {$\mu \to e \gamma$ in Theories With Dirac and
  Majorana Neutrino Mass Terms}},  {\em Phys. Rev. Lett.} {\bf 45} (1980) 1908.

\bibitem{Aubert:2009ag}
{\bf BaBar} Collaboration, B.~Aubert et~al., {\it {Searches for Lepton Flavor
  Violation in the Decays tau+- ---\ensuremath{>} e+- gamma and tau+-
  ---\ensuremath{>} mu+- gamma}},  {\em Phys. Rev. Lett.} {\bf 104} (2010)
  021802, [\href{http://arxiv.org/abs/0908.2381}{{\tt arXiv:0908.2381}}].

\bibitem{Adam:2013mnn}
{\bf MEG} Collaboration, J.~Adam et~al., {\it {New constraint on the existence
  of the $\mu^+ \to e^+\gamma$ decay}},  {\em Phys. Rev. Lett.} {\bf 110}
  (2013) 201801, [\href{http://arxiv.org/abs/1303.0754}{{\tt
  arXiv:1303.0754}}].

\bibitem{Baldini:2013ke}
A.~M. Baldini et~al., {\it {MEG Upgrade Proposal}},
  \href{http://arxiv.org/abs/1301.7225}{{\tt arXiv:1301.7225}}.

\bibitem{KATRIN:2021uub}
{\bf KATRIN} Collaboration, M.~Aker et~al., {\it {Direct neutrino-mass
  measurement with sub-electronvolt sensitivity}},  {\em Nature Phys.} {\bf 18}
  (2022), no.~2 160--166, [\href{http://arxiv.org/abs/2105.08533}{{\tt
  arXiv:2105.08533}}].

\end{thebibliography}\endgroup
\bibliographystyle{JHEP}
\end{document}